\documentclass[review]{elsarticle}

\usepackage{lineno,hyperref}
\modulolinenumbers[5]

\usepackage{afterpage}
\usepackage{amsmath}

\usepackage{amssymb}
\hypersetup{breaklinks=true} %VARIAS lineas
\usepackage[title]{appendix}% VICTOR
\usepackage{tikz} %VICTOR circulos
\newif\ifstartedinmathmode
\newcommand\encircle[1]{%
  \relax\ifmmode\startedinmathmodetrue\else\startedinmathmodefalse\fi%
  \tikz[baseline,anchor=base]{%
  \node[draw,circle,outer sep=0pt,inner sep=.2ex]
    {\ifstartedinmathmode$#1$\else#1\fi};}%
}
\newcommand{\refp}[1]{(\ref{#1})}
\usepackage{nomencl}
\makenomenclature
\usepackage{ifthen}
\renewcommand\nomgroup[1]{%
  \ifthenelse{\equal{#1}{A}}{%
    \item[\textbf{Acronyms}]}{%                A - Acronyms
  \ifthenelse{\equal{#1}{R}}{%
    \item[\textbf{Roman Symbols}]}{%           R - Roman
  \ifthenelse{\equal{#1}{G}}{%
    \item[\textbf{Greek Symbols}]}{%           G - Greek
  \ifthenelse{\equal{#1}{S}}{%
    \item[\textbf{Superscripts}]}{%            S - Superscripts
  \ifthenelse{\equal{#1}{U}}{%
    \item[\textbf{Subscripts}]}{%              U - Subscripts
  \ifthenelse{\equal{#1}{X}}{%
    \item[\textbf{Other Symbols}]}{%           X - Other Symbols
  {}}}}}}}}

\journal{Arxviv}

\begin{document}

\begin{frontmatter}

\title{Characteristic differential equation of a T-junction: diffusive shear work exchange from its head \emph{loss} coefficients.}

%\iffalse
\author[mymainaddress]{V\'ictor-Manuel Soto-Franc\'es\corref{mycorrespondingauthor}}
\cortext[mycorrespondingauthor]{Corresponding author}
\ead{vsoto@ter.upv.es}

\author[mymainaddress]{Jos\'e-Manuel Pinazo-Ojer}
\author[mymainaddress]{Emilio-Jos\'e Sarabia-Escriv\'a}
\author[umh]{Pedro-Juan Mart\'inez-Beltr\'an}

%\address{Dpto. Termodin\'amica Aplicada. Edificio 5J. ETSII. C/Camino de Vera S/n}



\address[mymainaddress]{Dpto. Termodin\'amica Aplicada. Edificio 5J. ETSII. C/Camino de Vera S/n. Universitat Polit\`ecnica de Val\`encia. 46022 Valencia, Spain}
\address[umh]{Universidad Miguel Hern\'andez, Dpto. Ingenier\'ia Mec\'anica y Energ\'ia, Elche, Spain}
%\address[mysecondaryaddress]{Universitat Polit\`ecnica de Val\`encia}
%\fi

\begin{abstract}
Negative head loss coefficients in branched junctions, has been controversial for long.
Herwig et al. showed, that the cause is a diffusive shear work exchange.
Based on their work, a new junction internal model is described, while the conventional head loss is named external.
The latter is obtained experimentally, while the first cannot.
This fact seems to push back reaching a practical solution.
Conventionally two head `loss' coefficients are required.
However the internal model needs three: two `pure' head loss coefficients and a work coefficient.
Based on previous works, the paper shows that the missing equation comes from the Minimum Energy Dissipation Principle (MinEDP).
The characteristic differential equation of a T-junction is discovered, which relates the two `pure' head loss coefficients.
A particular case is presented to illustrate how to obtain the internal model from the external one by using it.
The results are applied to empirical and numerical data from the same branched junction; Zhu's measurements (external) and Herwig's CFD computations (internal), respectively.
Finally, an example illustrates how powerful this new method is to analyse a new, recently published, exhaust return duct.
\end{abstract}

\begin{keyword}
negative loss coefficient\sep T-junction \sep energy dissipation \sep branching \sep MinEDP \sep Exhaust ventilation \sep Local drag coefficient
\end{keyword}

\end{frontmatter}

\nomenclature{$p$}{Pressure $[Pa]$}
\nomenclature{$\Delta p_T$}{Pressure drop of a network $[Pa]$}
\nomenclature{$z$}{Height $[m]$}
\nomenclature{$\bar{v}$}{Average cross-sectional velocity $[m/s]$}
\nomenclature{$g$}{Gravitational acceleration $[m/s^2]$}
\nomenclature{$\dot{L}$}{Loop (or pseudo-loop) flow rate $[m^3/s]$}
\nomenclature{$F$}{Global energy dissipation function per unit of reference volume flow $[Pa]$}
\nomenclature[G]{$\varphi$}{Energy dissipation per unit mass $[J/kg]$}
\nomenclature[U]{$R$,$r$}{Straight section}
\nomenclature[U]{$B$,$b$}{Side-branch section}
\nomenclature[U]{$C$,$c$}{Common section}
\nomenclature[S]{$\widehat{•}$}{Per unit of volume}
\nomenclature[S]{$\dag$}{False energy dissipation}
\nomenclature[G]{$\psi_j$}{Flow ratio $\dot{V}_j/\dot{V}_T$}
\nomenclature[G]{$\dot{\Phi}$}{Energy dissipation $[W]$}
\nomenclature[G]{$\alpha$}{Kinetic energy correction factor}
\nomenclature[G]{$\rho$}{Density $[kg/m^3]$}

%\linenumbers
\printnomenclature
%% FOREWORD
\section*{Foreword}
In order to fully understand the paper the reader needs to understand our three previous published papers  \cite{SOTOFRANCES2019181} \cite{HEFAT2019} and\cite{SOTO2021-MinEDP}.
We strongly encourage the reader to understand them.
They contain the ``old wine'' about steady-state analysis of flow networks in a new, clearer, bottle.

This paper has been uploaded to Arxiv, because after sending the paper to eight $Q1$ journals, only one editor sent it to some reviewers.
However, after one single response from one of them, unfortunately, the editor gave us no rebuttal opportunity to his/her comments.
Therefore before losing the results in a drawer and, while trying to publish a \emph{more applied or less mathematical} version of the previous papers for a better understanding by the engineering community, we have decided to leave the paper here for widespread revision, consideration and comments. 

%% SECTION  %%%%%%%%%%%%%%%%%%%%%%%%%%%%%%%%%%%%%%%%%%%%%%%%%%%%%%%%%%%%%%%%%%%%%%%%%%%%%%%%%%%%%%%%%%%%%
\section{Introduction}
\label{sec:introduccion}
The traditional head \emph{loss} coefficients at branched junctions may become negative.
This has been controversial for a long time (\cite{wood:1993} \cite{jamesliggett:1994} \cite{wood:1994}).
More recently Jaroslav Stigler \cite{jaroslavstigler:2006:1} published an interesting and critical paper about the traditional model of the T-junctions.
He argued :
\begin{quote}
They have not changed for more than 50 years. They have been derived on the base of unrealistic assumptions.
\end{quote}
Moreover he was clearly conscious about the associated problems:
\begin{quote}
They (the head loss coefficients at each branch) could be less than zero therefore they cannot be treated as loss coefficients.
\end{quote}
Stigler's alternative was to use external and physically meaningful magnitudes: a head \emph{loss} coefficient of the junction and the momentum coefficient $C_M$, which measures the force component exerted by the fluid over the junction in the direction of the common conduit.

However, in our opinion, only very recently, Herwig and Schmandt \cite{SCHMANDT2014191} \cite{SCHMANDT2015268} explained why the head \emph{loss} coefficients may become negative.
The reason is a diffusive shear work exchange between the streams.
However, this work interaction cannot be directly measured by empirical means.
Thus, in order to test their arguments, Herwig et al. performed several CFD runs on a convergent T-junction.
They argued that, what is traditionally called head \emph{loss} coefficients, should be renamed as head \emph{change} coefficients.
The reason is that the conventional coefficients contain both a \emph{pure} dissipative effect plus the aforementioned work interaction.
If the work received at one branch, from the other, overcomes the energy dissipation at that branch, then the work shows up as a negative head \emph{loss}.
Notice that even when both conventional head coefficients are positive, a work interaction may still exist, so to speak, `hidden' from our sight.
However, although based on sound physical principles according to Stigler's desire, the need for a CFD seemed to push back the finding of a practical solution to the `negative loss contradiction' of the current and useful, head \emph{change} coefficients.
\newline

In three previous works, the Minimum Energy Dissipation Principle (MinEDP) was used to find out the steady-state flow distribution of a flow network.
In \cite{SOTOFRANCES2019181} the MinEPD was applied to tree-shaped networks without work interaction at the junctions and the same year these results were extended to any type of network in \cite{HEFAT2019}.
Finally, in \cite{SOTO2021-MinEDP} the previous results were extended to any flow network with work interaction at its branched junctions.
In doing so, the Herwig's split of the head \emph{change} coefficients into two head \emph{loss} coefficients and a work coefficient was used to develop and prove that the new method, based on MinEDP, was equivalent to the energy balance.
So, easily a question arose: is it possible to use the \emph{two} conventional head \emph{change} coefficients to deduce the \emph{three} parameters needed by Herwig to model the branched junction?.
This paper proves its affirmative answer: the missing equation is based on the MinEDP.

%% SECTION %%%%%%%%%%%%%%%%%%%%%%%%%%%%%%%%%%%%%%%%%%%%%%%%%%%%%%%%%%%%%%%%%%%%%%%%%%%%%%%%%%%%%%%%%%%%%%
\section{Branched junction: external and internal models}
\begin{figure}[ht]
	\centering
	\includegraphics[width=0.95\textwidth]%
	{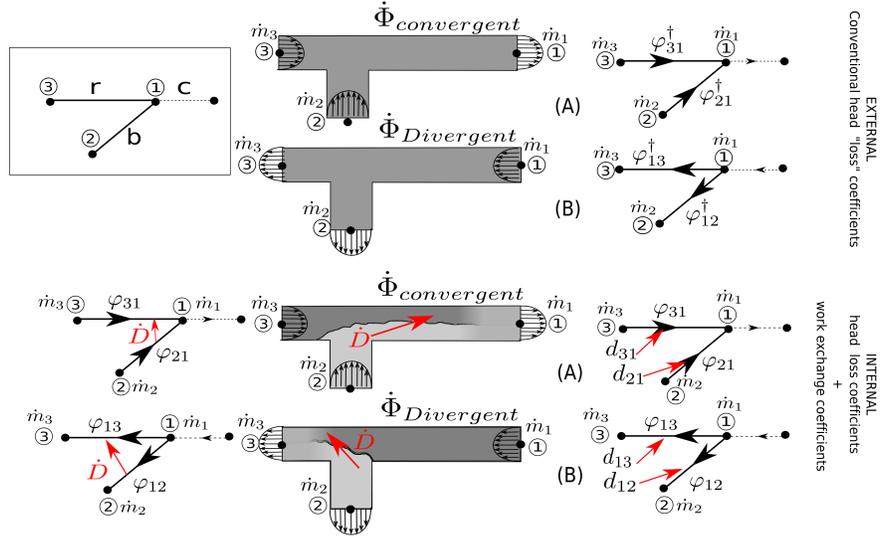}% picture filename
	\caption{External (top) and Internal (bottom) models of a branched junction: (A) Converging T-junction (B) Diverging T-junction. $1$ common port, $3$ main port, $2$ side-branch port. }
	\label{fig:external-internal}
\end{figure}

The figure \refp{fig:external-internal} shows the scheme of two models of a branched junction.
Regardless of the flow sense (convergent or divergent), the section connecting port \encircle{3} to port \encircle{1} is the straight section, denoted by $r$, while the section connecting \encircle{2} to \encircle{1} is called the side-branch section, denoted by $b$.
Finally, the incoming or outgoing flow section is called the common section and denoted by $c$.
%% FIG %%
The one at the top is the conventional model based on the head \emph{loss} coefficient at each branch.
As aforementioned in \S\ref{sec:introduccion}, head \emph{change} coefficient is the correct name and will be used from now on.
The symbols $\widehat{\varphi}^{\dag}_{ij}$ represent the head \emph{change} from the $i$-port to the $j$-port due to the presence of the junction.
They are usually represented by $\Delta p_{ij}$ instead of $\widehat{\varphi}^{\dag}_{ij}$.
Sometimes they are just tabulated as $\Delta p_r$ and $\Delta p_b$ for the straight and side branches respectively (see \cite{idelchik1986}).
They are considered, conventionally and mistakenly, as a \emph{loss} (thus their traditional name follows).
According to Herwig's nomenclature, the symbol $\varphi$ will be used for the actual energy dissipation and that is why, in order to remark the wrong interpretation of $\Delta p_{ij}$, the symbol $\widehat{\varphi}^{\dag}_{ij}$ with a $\dag$, is used here to remember that it is a \emph{false loss}.
These head \emph{changes} are fitted to a potential law, i.e, to the square of the flow mean velocity at the common branch.
The expressions used are:
$$ \widehat{\varphi}^{\dag}_r= C_r \cdot \frac{\rho \bar{v}^2_c}{2}$$
$$ \widehat{\varphi}^{\dag}_b= C_b \cdot \frac{\rho \bar{v}^2_c}{2}$$
Obviously, $C_r$ and $C_b$ are called the head \emph{change} (formerly \emph{loss}) coefficients.
The energy loss or energy dissipation of the junction can be written as:
$$ \dot{\Phi}_{junction}=\overbrace{C_{junction}\cdot \frac{\rho \bar{v}^2_1}{2}}^{ \widehat{\varphi}_{junction}}\cdot \dot{V}_1=
   \widehat{\varphi}^{\dag}_r\cdot \dot{V}_r+\widehat{\varphi}^{\dag}_b\cdot \dot{V}_b=
\left( C_r\cdot \dot{V}_r+C_b\cdot \dot{V}_b\right)\cdot \frac{\rho \bar{v}^2_1}{2}$$
Notice that $\widehat{\varphi}_{junction}$ has no $\dag$, since it is always a \emph{pure} dissipation or a \emph{true} loss.
Additionally, by calling $x$ the side-branch ratio of flow rates, i.e., $x=\psi_2=\dot{m}_2/\dot{m}_1=\dot{V}_{2\equiv b}/\dot{V}_{1\equiv c}$, the following relationship holds:
%% EQ %%
\begin{equation}
C_{junction}=C_r\cdot \frac{\dot{V}_r}{\dot{V}_c} +C_b\cdot \frac{\dot{V}_b}{\dot{V}_c}=C_r\cdot(1-x)+C_b\cdot x
\label{ec:Cjunction}
\end{equation}
The head \emph{change} coefficients $C_r$ and $C_b$ are not constant, in general, since they depend (for branched junctions), on the flow ratio (for a detailed discussion see \cite{SCHMANDT2014191}).
%% EQ %%
\begin{equation}
\begin{split}
 \varphi^{\dag}_r(x)=& C_r(x) \cdot \frac{\rho \bar{v}^2_c}{2} \\
 \varphi^{\dag}_b(x)=& C_b(x) \cdot \frac{\rho \bar{v}^2_c}{2}
 \end{split}
 \label{ec:CrCb}
\end{equation}
The conventional model will also be called here, the \emph{external} model, because it can be obtained directly by measuring the junction in a laboratory set-up.

At the bottom of figure \refp{fig:external-internal} a new model based on Herwig's paper \cite{SCHMANDT2015268} is shown.
In opposition to the \emph{external} model this will be called \emph{internal} because it cannot be measured directly in the laboratory.
The new model is characterised by two types of terms: a \emph{pure} energy dissipation or loss $\varphi$ and a work exchange $d$.
The rate of work exchange is computed by \cite{SCHMANDT2015268}:
%% EQ %%
\begin{equation}
      \dot{D}=\int_A{\overline{\overline \tau}\cdot \vec{v}\cdot d\vec{A}}
      \label{ec:D}
\end{equation}
The meaning of $\dot{D}$ is the total diffusive work transfer between the two streams as a consequence of the shear stress tensor $\overline{\overline \tau}$.
In volumetric form eq. \refp{ec:D} is written $\dot{D}'''=\nabla \cdot (\overline{\overline \tau}\cdot \overline{v})$ as the local stress work rate and, finally, its specific value per unit of mass is called $d=\dot{D}/\dot{m}=\int{\dot{D}'''\cdot dV}/\dot{m}$ (see \cite{SCHMANDT2015268} for the details).
If $d$ is referred to the flow rate at each branch then there are two specific values $d_r$ and $d_b$, for the same work interaction $\dot{D}$.
Nevertheless, they are useful, since it allows us to write the extended Bernoulli's equation or energy change at the convergent junction, for instance, as:
%% EQ %%
\begin{equation}
	\begin{split}
		e_{m,3}-e_{m,1}=\frac{p_{3}}{\rho}-\frac{p_1}{\rho}+\frac{\alpha_3\bar{v}^2_3}{2}-\frac{\alpha_1\bar{v}^2_1}{2}+gz_3-gz_1=& \varphi_{31}-d_{31}=\varphi_{31}^{\dag} \\
		e_{m,2}-e_{m,1}=\frac{p_{2}}{\rho}-\frac{p_1}{\rho}+\frac{\alpha_2\bar{v}^2_2}{2}-\frac{\alpha_1\bar{v}^2_1}{2}+gz_2-gz_1=& \varphi_{21}-d_{21}=\varphi_{21}^{\dag}
	\end{split}
\end{equation}
where $e_{m,i}$ is the specific mechanical energy per unit mass at the $i$-node.
Therefore the relationship between the \emph{false} energy loss $\varphi^{\dag}$ and the actual one $\varphi$ is :
%% EQ %%
\begin{equation}
\varphi^{\dagger}=\varphi-d \quad \text{ or } \quad \widehat{\varphi}^{\dagger}=\widehat{\varphi}-\widehat{d}
\label{ec:varphi-daga}
\end{equation}
Notice that $ d_r\cdot \dot{m}_r+d_b\cdot \dot{m}_b=0$ or in volumetric form $\widehat{d}_r\cdot \dot{V}_r+\widehat{d}_b\cdot \dot{V}_b=0$, i.e., the work received by one branch equals the work delivered by the other.
Therefore, in fact, the only independent parameters of the \emph{internal} model are three: $\{\varphi_r,\varphi_b,d\}$.
Similarly to the conventional $\widehat{\varphi}^{\dag}$ (see \cite{SOTO2021-MinEDP}), the following equations can be written:
\begin{equation}
\begin{split}
	\widehat{\varphi}_r&=C_{\varphi,r}\cdot \frac{\rho \bar{v}^2_c}{2} \quad,\quad \widehat{\varphi}_b= C_{\varphi,b}\cdot \frac{\rho \bar{v}	^2_c}{2} \\
	\widehat{d}_r&=C_{d,r}\cdot \frac{\rho \bar{v}^2_c}{2} \quad,\quad \widehat{d}_b= C_{d,b}\cdot \frac{\rho \bar{v}^2_c}{2} \quad , \quad
	\widehat{d}  =C_{\bar{d}}\cdot \frac{\rho \bar{v}^2_c}{2}
\end{split}
\end{equation}
Therefore, according to eqs. \refp{ec:varphi-daga} and \refp{ec:CrCb}, the relationships among the old and the new coefficients are:
%% EQ %%
\begin{equation}
\begin{split}
C_r(x) = C_{\varphi,r}(x)-C_{d,r}(x) \\
C_b(x) = C_{\varphi,b}(x)-C_{d,b}(x)
\end{split}
\label{ec:C-C}
\end{equation}
thus, substituting the previous  eq.\refp{ec:C-C} into eq. \refp{ec:Cjunction}, $C_{junction}$ for the \emph{internal} model is written:
%% EQ %%
\begin{equation}
C_{junction}=C_{\varphi,r} \cdot \frac{\dot{V}_r}{\dot{V}_c} +C_{\varphi,b}\cdot \frac{\dot{V}_b}{\dot{V}_c}=C_{\varphi,r}\cdot(1-x)+C_{\varphi,b}\cdot x
\label{ec:Cjunction-intern}
\end{equation}
since $C_{d,r}\cdot (1-x)+C_{d,b}\cdot x=0$.
For more details about the \emph{external} and \emph{internal} coefficients the reader is referred to \cite{SOTO2021-MinEDP}.
\newline

At first sight, it is seemingly impossible to obtain $\{\varphi_r,\varphi_b,d\}$ from $\{\varphi_r^{\dag},\varphi_b^{\dag}\}$, since we have only the two equations \refp{ec:C-C}.
In the next sections, it will be shown how the MinEDP provides the missing equation which allows the transformation of one model into the other.
The \S\ref{sec:MinEDP} presents only the key conclusions from \cite{SOTOFRANCES2019181} \cite{HEFAT2019} \cite{SOTO2021-MinEDP}, needed here.
Before going into the mathematical details, \S\ref{sec:Qualitative} is employed to describe qualitatively the physical implications of the following sections.
The mathematical equivalent of \S\ref{sec:Qualitative} is found in \S\ref{sec:characteristic}.
Within this latter section an unexpected universal characteristic differential equation is discovered.
It relates $\varphi_r$ and $\varphi_b$ for any branched junction (with three ports).
In \S\ref{sec:particular} the previous universal relationship is applied to a particular case.
Finally, in \S\ref{sec:validation} the previous particular case is applied to a concrete and real case of a convergent T-junction measured by Zhu \cite{ZhuThesis}.
Notice that the measurements correspond, necessarily, to the \emph{external} model or conventional head \emph{change} coefficients.
Our MinEDP method is compared with Zhu's experimental results  but also, with a paper that estimates the internal work exchange $d$, (see \cite{SCHMANDT2015268}).
Fortunately, this latter paper uses also the same Zhu's results for validating their CFD outcomes and therefore it allows us to `see' or to get an estimation of what is going on \emph{inside} the Zhu's results.

%% SECTION %%%%%%%%%%%%%%%%%%%%%%%%%%%%%%%%%%%%%%%%%%%%%%%%%%%%%%%%%%%%%%%%%%%%%%%%%%%%%%%%%%%%%%%%%%%%%%%%%
\section{Brief review of MinEDP}
\label{sec:MinEDP}
The aim of this section is just to provide some new key ideas developed in our previous works \cite{SOTOFRANCES2019181} \cite{HEFAT2019} \cite{SOTO2021-MinEDP}.
For any flow network, by fixing beforehand one or several flows, its global energy dissipation under such imposed flow rate/s, can be expressed as:
%% EQ %%
\begin{equation}
(F\circ h \circ g) (x_1,\dots,x_n)=\frac{\dot{\Phi}}{\dot{V}_T}=\widehat{\varphi}_T=\Delta p_T
\label{ec:disipacion-compo}
\end{equation}
where, $x_i$ are the $n$ independent loop (or pseudo-loop) flow ratios of the network referred to a reference flow rate $\dot{V}_T$, i.e., $x_i=\dot{L}_i/\dot{V}_T$.
In fact, $\widehat{\varphi}_T$ is computed as the composition of three functions $F$, $h$ and $g$.
The first one, $g$ is :
%% EQ %%
\begin{equation}
g:(x_1,\dots,x_n)\rightarrow (\psi_1,\psi_2,\dots,\psi_{nsect})
\end{equation}
, in concrete, is a linear map from the independent flow ratio vector $\vec{x}$ to the vector $\vec{\psi}$ of the flow ratios at each of the $nsect$ sections of the network.
The $h$ map takes the absolute value of each component of $\vec{\psi}$:
%% EQ %%
\begin{equation}
h:(\psi_1,\psi_2,\dots,\psi_{nsect})\rightarrow (|\psi_1|,|\psi_2|,\dots,|\psi_{nsect}|)
\end{equation}
and $F(h(\vec{\psi}))$ has the form:
%% EQ %%
\begin{equation}
(F\circ h)(\vec{\psi})=\frac{\dot{\Phi}}{\dot{V}_T}=\sum^{j=nsect}_{j=1} {[\widehat{\varphi}^{\dag}]_j\cdot |\psi_j|}
\label{ec:disipacion-red}
\end{equation}
where by $[\widehat{\varphi}^{\dag}]_j$ it is meant the sum of all the head \emph{changes}, due to all components in series, within the $j$-section, including the branch of a junction belonging to that $j$-section.
Finally:
%% EQ %%
\begin{equation}
\widehat{\varphi}^{\dag}=\widehat{K}^{\dag}\cdot |\psi|^{m}
\end{equation}
where the \emph{energy} factor $\widehat{K}^{\dag}$ is related to the head \emph{change} coefficients of the component and $m$ is an exponent not necessarily equal to $2$ (see the details in\cite{SOTOFRANCES2019181}).
So, eq.\refp{ec:disipacion-red} also admits a $K$-form as follows:
\begin{equation} 
(F\circ h)(\vec{\psi})=\frac{\dot{\Phi}}{\dot{V}_T}=\sum^{j=nsect}_{j=1} {[\widehat{K}^{\dag}]_j\cdot |\psi_j|^{(m+1)}}
\label{ec:disipacion-red-K}
\end{equation}
Equation \ref{ec:disipacion-compo} or \ref{ec:disipacion-red-K} can be written just as $F(x_1,\dots,x_n)$ and represents the global dissipation per unit of reference flow.
In \cite{SOTOFRANCES2019181} it was shown that if there are no work exchanges at the branched junctions, i.e., all components are \emph{purely} dissipative ones, then $[\widehat{K}^{\dag}]$ is in fact $[\widehat{K}]$ (i.e., drops the $\dag$) and the steady-state solution of the network is found at the minimum of eq.\refp{ec:disipacion-red-K}.
However, if the work interaction at the branched junctions is considered, then the steady-state of the network is not at the minimum of eq.\refp{ec:disipacion-red-K}, although it tries to get as close as possible to that minimum.
Stated in other words, finding the steady-state of the network becomes solving a constrained minimization problem, due to the internal work exchange (see \cite{SOTO2021-MinEDP} for the details).
\newline

This difference between the properties of the steady-state solution related with the global dissipation function of the network and the existence, or not, of a work exchange at the junctions is crucial for our purposes, as it will be shown qualitatively in the next section.

%% SECTION %%%%%%%%%%%%%%%%%%%%%%%%%%%%%%%%%%%%%%%%%%%%%%%%%%%%%%%%%%%%%%%%%%%%%%%%%%%%%%%%%%%%%%%%%%%%%%%%%%%%%%%%%%5
\section{Estimation of the work exchange}
\label{sec:Qualitative}
The aim of this section is to describe, qualitatively, how the specific work exchange $d$ inside a branched junction and in steady-state flow, can be obtained by using the \emph{external} knowledge provided by the head \emph{change} coefficients and the MinEDP.

\begin{figure}[ht]
	\centering
	\includegraphics[width=1.0\textwidth]%
	{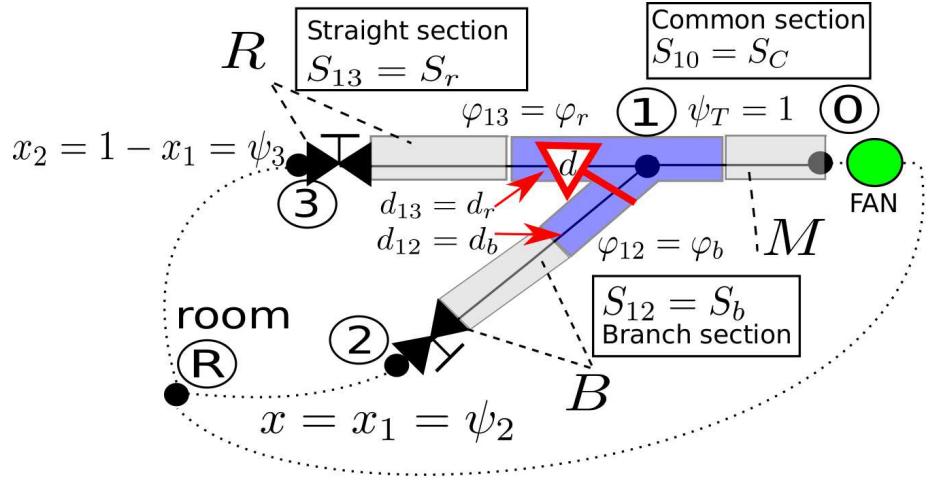}% picture filename
	\caption{Scheme of a thought experimental set-up.The names correspond to a divergent junction. In the case of a convergent junction the sub-script numbers of the \emph{internal} model reverse their order; $d_{31}$, $d_{21}$, $\varphi_{31}$ and $\varphi_{21}$. }
	\label{fig:exampleD}
\end{figure}

The figure \refp{fig:exampleD} describes a thought experimental set-up.
It has three sections: the straight section $S_{13}\equiv S_r$, the common or main section $S_{10}\equiv S_C$ and the side-branch section $S_{12}\equiv S_b$.
The straight and side-branch sections have valves which allow us to adjust the flow rate ratio $x$ through the side-branch.
The main section has an impeller and the set-up can work in supply and return modes, i.e., the junction can be divergent or convergent, respectively.
The four parameters of the \emph{internal} model of the junction are named as shown in the figure \refp{fig:exampleD}: $\varphi_r$ and $\varphi_b$ are the \emph{pure} dissipation per unit mass flow rate at the straight and side-branch respectively and similarly, $d_r$, $d_b$ are used for the exchanged diffusive shear work between the streams.
Figure \refp{fig:exampleD} shows the names for the supply mode, however for our purposes, since the sense does not matter to the exposition, $\varphi_{12}$ and $\varphi_{21}$ will be just named $\varphi_b$ and analogously, with the other three parameters.
%% FIG %%
%%  %%
Without loss of generality, the measured loss coefficient of the junction $C_{junction}$ and the head \emph{change} at one branch ${C}_{31}={C}_r$ or ${C}_{21}={C}_b$ are going to be used as inputs.
Therefore, the other head \emph{change} will be an output.

Let's imagine that the junction is tested keeping the common flow rate $\dot{V}_1\equiv \dot{V}_c$ constant.
Therefore, the flow distribution $x$ is the only independent variable.
As mentioned, the steady-state value of $x$ can be adjusted arbitrarily with the valves.
Once a full empirical test of the junction is made, the actual head \emph{change} coefficients $C_b(x)$ and $C_r(x)$ (\emph{external} model) are found.
This is roughly, the actual procedure to measure these coefficients at the laboratory.
Recall that the loss coefficient of the junction can be computed as: $C_{junction}(x)=C_b(x)\cdot x + C_r(x)\cdot (1-x)$.

\begin{figure}[ht]
	\centering
	\includegraphics[width=0.75\textwidth]%
	{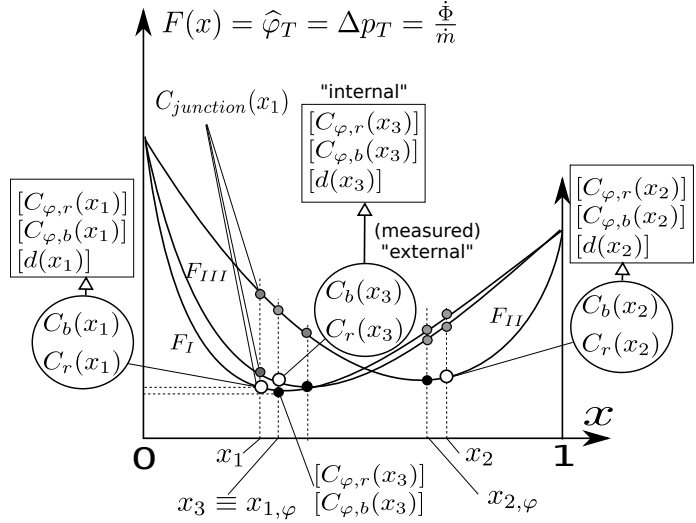}% picture filename
	\caption{Qualitative representation of the dissipation function for various sets of valve position ($(I)$,$(II)$ and $(III)$). Full circles are \emph{artificial} points while void circles are real, measurable, working points.
	All points (artificial or real) in the same vertical share the same values $\widehat{C}_{b\varphi}$,$\widehat{C}_{r,\varphi}$ and $\widehat{D}$ since the flow ratio is the same.}
	\label{fig:exampleD-2}
\end{figure}

Figure \refp{fig:exampleD-2} shows, qualitatively, the specific energy dissipation function $F(x)$ for three different position of the valves named $F_I$, $F_{II}$ and $F_{III}$.
Let's state, more clearly, how these curves are obtained.
Once the valves are fixed at a certain position, let's say case $I$, the graph of $F_I$ would be obtained by `virtually' varying the flow ratio $x$ from $0$ to $1$, i.e., without changing the position of the valves, and computing at each $x$ the total energy dissipation due to all the components: the straight conduits, the valves and the junction itself.
This can be done because the $C_{junction}(x)$ function, is already known from the previous experimental tests.
As it was shown in \cite{SOTO2021-MinEDP}, on these curves the \emph{external} work is zero, since the \emph{internal} work received by both streams cancels each other out.
In other words, the global dissipation $F$ is not aware of any work exchange.
In papers \cite{SOTOFRANCES2019181} and \cite{SOTO2021-MinEDP}, it was proven that the actual steady-state coincides with the minimum of the energy dissipation function, whenever there is no internal work interaction $d=0$, but if $d\neq 0$ then it is not a minimum.
In figure \refp{fig:exampleD-2} there are three types of points represented by full, grey and void dots.
Full and grey dots are `artificial' while void dots are `real' (measurable at the lab) operating points of the set-up of figure \refp{fig:exampleD}.
A full dot is a particular case of a grey dot on a certain dissipation curve, since it corresponds to the minimum dissipation of a certain function $F$.
For instance, $F_I$ has a minimum at $x_{1,\varphi}$, i.e., $min_{x}{F_I(x)}=F_I(x_{1,\varphi})$.
On the same curve there is a void dot with $x=x_1$ which represents the actual operating (achievable or measurable) point for case $I$.
Notice that for a point on any $F$-curve with the same $x'$ coordinate, i.e., in the same vertical line, the junction contribution to the energy dissipation, is the same in all the $F$-curves.
See for instance $x_1$ in fig.\refp{fig:exampleD}.
The $F_I(x_1)$ is an actual measurement, a void dot, and the other two are \emph{fictitious} or \emph{artificial} values, the grey dots given by $F_{II}(x_1)$ and $F_{III}(x_1)$, which correspond to position $II$ and $III$  of the valves.
All three points have the same value $C_{junction}(x_1)$.
%% FIG %%
Now, let's make a \emph{fictitious} test.
Let's us assume that actual energy dissipation coefficient $C_{junction}(x)$ for each flow distribution $x$ is valid.
However, let's remove \emph{artificially} the work exchange $d$.
According to \cite{SOTOFRANCES2019181} \cite{SOTO2021-MinEDP}, the flow distribution, for the valves in position $I$, coincides with the minimum of the energy dissipation function $F_I$ at $x_{1,\varphi}$.
The point $x_{1,\varphi}$ is \emph{artificial} since for case $I$ the actual measurable point is $x_1$.
However the key point of the previous action is that it allows us to split the \emph{pure} energy losses between the two branches at $x_{1,\varphi}$, that is, allows us to compute the internal head \emph{loss} coefficients: $C_{\varphi,b}(x_{1,\varphi})$, $C_{\varphi,r}(x_{1,\varphi})$.
More exactly, as next \S\ref{sec:characteristic} will show, a functional relationship between both pure dissipations.
Otherwise stated, since the global energy dissipation function by itself, does not contain the work internally exchanged, it can be used to establish a relationship between the \emph{internal} energy dissipation at each branch based on the actual, measured, dissipation.
Notice that the energy dissipation split is not found at $x_1$ but at $x_{1,\varphi}$ which in turn, corresponds to the \emph{real} steady-state $x_3$, only measurable when the valves are at the position $III$.
Nevertheless, the \emph{artificial} point $x_{1,\varphi}$ and the \emph{real} one $x_3$ share the same vertical line and therefore, the junction is dissipating energy at the same rate at both points.
This fact can be expressed by equation \refp{ec:junction-x1}:
%% EQ %%
\begin{equation}
\begin{split}
	{C}_{junction}(x_3)=&{C}_{b}(x_3)\cdot x_3+{C}_{r}(x_3)\cdot(1-x_3)=\\
	                   =&{C}_{\varphi,b}(x_{1,\varphi})\cdot x_{1,\varphi}+{C}_{\varphi,r}(x_{1,\varphi})\cdot(1-x_{1,\varphi})=\\
	                   =&{C}_{\varphi,b}(x_3)\cdot x_3+{C}_{\varphi,r}(x_3)\cdot(1-x_3)
\end{split}
	\label{ec:junction-x1}
\end{equation}
However the head \emph{change} coefficients measured at $x_3$, $C_{b}(x_3)$ and ${C}_{r}(x_3)$, are not equal to ${C}_{\varphi,b}(x_3)$ and ${C}_{\varphi,r}(x_3)$ estimated previously by our \emph{artificial} test, since that test ignores (removes) the work exchange.
Precisely, this difference leads to find out the value $d(x_3)$ or equivalently the work coefficient $C_{\bar{d}}$.
Notice that for this last step, only one of the two coefficients $C_{b}(x_3)$ or ${C}_{r}(x_3)$ is needed.
The other will become an output of the method and must be equal to the measured value.

Proceeding in the same way for another position of the valves, let's say $II$, the internal values: $C_{\varphi,b}(x_{2,\varphi})$, $C_{\varphi,r}(x_{2,\varphi})$
would be obtained corresponding to another \emph{real} measurement with the valves at a different position (not represented in figure \refp{fig:exampleD-2}.
Therefore two important sets of points are obtained: on one hand the \emph{real} or measured set represented by the void dots and the \emph{artificial} or fictitious one represented by the full dots and computed based on the MinEDP.
The first contains, at each point, the \emph{external} coefficients while the second contains only the virtual \emph{internal pure} dissipation coefficients.

In summary, by using two `pieces'  of \emph{external} and measurable information, three `pieces' of \emph{internal} information can be obtained \{ ${C}_{\varphi,b}(x)$ , ${C}_{\varphi,r}(x)$, $d(x)$ \}.
The input data can be: either $C_{junction}(x)$ and the head \emph{change} at any of the two branches, or the two head \emph{change} coefficients $C_b(x)$ and $C_r(x)$ (since by eq. \refp{ec:Cjunction} $C_{junction}(x)$ can be obtained).
The equation that closes the problem derives from the application of the MinEDP.

Next section clarifies the aforementioned qualitative ideas.

%% SECTION $$
\section{Characteristic differential equation of a branched junction.}
\label{sec:characteristic}
This section develops, mathematically, the previous ideas about how to determine the \emph{internal} parameters of the branched junction model from \emph{external} measurements.

Let's go back to our test set-up of figure \refp{fig:exampleD}.
By making the conventional energy balance (Bernoulli's equation), the head loss along the path \encircle{1}-\encircle{3}-\encircle{R} must be equal to the loss along \encircle{1}-\encircle{2}-\encircle{R}, since they form a `pseudo-loop'.
As was proved in \cite{SOTOFRANCES2019181}, without loss of generality, let's assume that in the branches the head loss can be expressed as $R\cdot \rho\bar{v}^2_r/2$ and $B\cdot \rho\bar{v}^2_b/2$, where $R$ and $B$ are the loss coefficients which include the straight conduits and the valves.
In order to simplify the mathematics, $R$ and $B$ remain constant.
Therefore the energy balance can be written as:
%% EQ %%
\begin{equation}
	{C}_b\cdot\rho\frac{\bar{v}^2_c}{2}+{B}\cdot\rho\frac{\bar{v}^2_b}{2}-\left( {C}_r\cdot\rho\frac{\bar{v}^2_c}{2}+{R}\cdot\rho\frac{\bar{v}^2_r}{2}\right)=0
\end{equation}
or using  $\dot{V}_b=A_b\cdot \bar{v}_b$ and $x=\dot{V}_b/\dot{V}_T$:
%% EQ %%
\begin{equation}
	{C}_b+{B}\left(\frac{A_c}{A_b}\right)^2\cdot x^2-\left( {C}_r+{R}\left(\frac{A_c}{A_r}\right)^2\cdot (1-x)^2\right)=0
	\label{ec:balance-ejemploD}
\end{equation}
Let's make explicit the influence of the work exchange in the previous expression \refp{ec:balance-ejemploD}.
In order to do that, first, some new expressions must be deduced for the \emph{internal} model in order to express the work exchange as a single parameter.
The expression for $\varphi^{\dagger}$ (see eq. \refp{ec:varphi-daga}) allows us to write:
%% EQ %%
\begin{equation}
	\begin{split}
	\varphi^{\dagger}_b&=C_b\frac{\bar{v}_c^2}{2}=C_{\varphi,b}\frac{\bar{v}^2_c}{2}-C_{d,b}\frac{\bar{v}^2_c}{2}\\
	\varphi^{\dagger}_r&=C_r\frac{\bar{v}_c^2}{2}=C_{\varphi,r}\frac{\bar{v}^2_c}{2}-C_{d,r}\frac{\bar{v}^2_c}{2}
	\end{split}
	\label{ec:varphi-dagger-1}
\end{equation}
%%
%% EQ %%
For the case of a junction:
\begin{equation}
	\begin{split}
		\dot{m}_{r}\cdot d_{r}=& -\dot{m}_{b}\cdot d_{b}=\dot{D}=\dot{V}_r\cdot \widehat{d}_{r}=-\dot{V}_b\cdot \widehat{d}_{b}\\
		d_r\cdot \dot{m}_r+d_b\cdot \dot{m}_b=&\widehat{d}_{r}\cdot \dot{V}_r+\widehat{d}_{b}\cdot \dot{V}_b=0
	\end{split}
	\label{ec:DWork}
\end{equation}
Combining eq.\refp{ec:DWork} and eq.\refp{ec:varphi-dagger-1}:
\begin{equation}
	\begin{split}
		\dot{D}&=-\dot{m}_b\cdot d_{b}=-\frac{\dot{m}_b}{\dot{m}_c}\cdot d_{b}\cdot \dot{m}_c=-x\cdot d_{b}\cdot \dot{m}_c\\
	          d&=\frac{\dot{D}}{\dot	{m}_c}=-x\cdot C_{d,b}\cdot \frac{\bar{v}^2_c}{2}\\
	\frac{d}{x}&=-C_{d,b}\cdot\frac{\bar{v}^2_c}{2}
	\label{ec:d-x}
\end{split}
\end{equation}
and analogously $d/(1-x)=C_{d,r}\cdot\frac{\bar{v}^2_c}{2}$.
We can define a new parameter $C_{\bar{d}}$ as $C_{\bar{d}}=d/\frac{\bar{v}^2_c}{2}$.
It is a work exchange coefficient and represents the work transferred from the side-branch to the straight branch.
In this way, it will be avoided the use of two coefficients for the work.
Using eqs.\refp{ec:varphi-dagger-1} and \refp{ec:d-x}, it can be written:
%% EQ %%
\begin{equation}
	\begin{split}
		{C}_b&={C}_{b,\varphi}-{C}_{b,d}={C}_{b,\varphi}+\frac{C_{\bar{d}}}{x}\\
		{C}_r&={C}_{r,\varphi}-{C}_{r,d}={C}_{r,\varphi}-\frac{C_{\bar{d}}}{(1-x)}
	\end{split}
	\label{ec:C-ejemploD}
\end{equation}
Additionally, the following also holds:
%% EQ %%
\begin{equation}
	\underbrace{{C}_{b,d}\cdot x}_{-C_{\bar{d}}}+\underbrace{{C}_{r,d}\cdot (1-x)}_{C_{\bar{d}}}=0
	\label{ec:Cdb+Crd=0}
\end{equation}
So, now, going back to equation \refp{ec:balance-ejemploD}, the conventional energy balance, but written with the \emph{internal} model, has the form:
%% EQ %%
\begin{equation}
	{C}_{b,\varphi}+{B}\left(\frac{A_c}{A_b}\right)^2\cdot x^2-\left( {C}_{r,\varphi}+{R}\left(\frac{A_c}{A_r}\right)^2\cdot (1-x)^2 \right)=
	-\left( \frac{C_{\bar{d}}}{x}+\frac{C_{\bar{d}}}{(1-x)} \right)
	\label{ec:balance-ejemploD-final}
\end{equation}
On the other side, by assuming as before, that the loss coefficient $M$, at the common section (see fig. \refp{fig:exampleD}), remains constant, then the energy dissipation of our set-up $\dot{\Phi}$, has the following expression: 
%% EQ %%
\begin{equation}
	\dot{\Phi}=
	\left({M}\frac{\rho\bar{v}^2_c}{2}\right) \cdot \dot{V}_c+
	\left({C}_{b}\frac{\rho\bar{v}^2_c}{2}\right) \cdot \dot{V}_b+\left({B}\frac{\rho\bar{v}^2_b}{2}\right) \cdot \dot{V}_b+
	\left({C}_{r}\frac{\rho\bar{v}^2_c}{2}\right) \cdot \dot{V}_r+\left({R}\frac{\rho\bar{v}^2_r}{2}\right) \cdot \dot{V}_r
	\label{ec:disipacion-ejemploD-0}
\end{equation}
based on using the head \emph{change} coefficients (external model).
If this same eq. \refp{ec:disipacion-ejemploD-0} is rewritten with the \emph{internal} model then:
%% EQ %%
\begin{equation}
	\dot{\Phi}(x)=\frac{\rho\dot{V}^3_c}{2A_c^2}
	\left[
		{M}+{C}_{\varphi,b}\cdot x+
		{B}\left(\frac{A_c}{A_b}\right)^2\cdot x^3+
		{C}_{\varphi,r}\cdot(1-x)+{R}\left( \frac{A_c}{A_r}\right)^2\cdot (1-x)^3
	\right]
	\label{ec:disipacion-ejemploD}
\end{equation}
As was expected, eq. \refp{ec:disipacion-ejemploD} ignores the existence of $C_{\bar{d}}$.
According to the exposition in the previous \S\ref{sec:Qualitative}, the derivative of eq. \refp{ec:disipacion-ejemploD} must be zero at an \emph{artificial} point.
It is the stationarity condition of $\dot{\Phi}$ when there is no work exchange $d=0$.
%% EQ %%
\begin{equation}
	\frac{d\dot{\Phi}}{dx}=
	\frac{d{C}_{\varphi,b}}{dx}\cdot x+{C}_{\varphi,b}+ 3{B}\left(\frac{A_c}{A_b}\right)^2\cdot x^2+
	\frac{d{C}_{\varphi,r}}{dx}\cdot(1-x)-{C}_{\varphi,r}
	-3{R}\left(\frac{A_c}{A_r}\right)^2\cdot (1-x)=0
\end{equation}
By adding and subtracting these two terms $2{C}_{\varphi,b}$ y $2{C}_{\varphi,r}$ and reorganizing, the previous equation leads to:
%% EQ %%
\begin{equation}
	\begin{split}
		& 3\cdot\left[ {C}_{\varphi,b}+{B}\left(\frac{A_c}{A_b}\right)^2-\left( {C}_{\varphi,r}+{R}\left(\frac{A_c}{A_r}\right)^2\cdot (1-x)^2\right) \right] \\
	& +\left[
		-2{C}_{\varphi,b}
	+\frac{d{C}_{\varphi,b}}{dx}\cdot x
	+2{C}_{\varphi,r}
	+\frac{d{C}_{\varphi,r}}{dx}\cdot(1-x)
	\right]=0
	\end{split}
	\label{ec:derivada-ejemploD}
\end{equation}
The first term, between square brackets in eq.\refp{ec:derivada-ejemploD} equals the left hand side of the energy balance eq.\refp{ec:balance-ejemploD-final}.
Since the \emph{artificial} test assumes that there is no internal work exchange, i.e., $d=0$ or $C_{\bar{d}}=0$ then that first term must be zero.
Note that this is a forced condition, according to our fictitious test, as a way to remove the work.

Therefore the second term inside the square bracket must be also zero at the \emph{artificial} or fictitious steady-state (represented by a full dot in the previous section \S\ref{sec:Qualitative}).
\emph{Finally, in this way, we arrive at an important, and somehow surprising or unexpected, differential equation that establishes a relationship between the two head \texttt{loss} coefficients $C_{\varphi,b}$ and $C_{\varphi,r}$, of the \texttt{internal} model}.
%% EQ %%
\begin{equation}
	-2{C}_{\varphi,b}
	+\frac{d{C}_{\varphi,b}}{dx}\cdot x
	+2{C}_{\varphi,r}
	+\frac{d{C}_{\varphi,r}}{dx}\cdot(1-x)=0
	\label{ec:relacion-ejemploD}
\end{equation}

\section{Particular case: $C_{\varphi,r}$ is a $2^{nd}$ order polynomial}
\label{sec:particular}
This section illustrates the implications of the general characteristic equation \refp{ec:relacion-ejemploD}.

The starting data are \{${C}_{junction}$, ${C}_{r}$\} or equivalently \{$C_{b}$, $C_r$\}, previously measured at the laboratory.
Let's assume, as a particular case, that the head \emph{change} coefficients can be fitted to a $2^{nd}$ order polynomial.
Therefore, the head \emph{loss} coefficient at the side-branch has the form:
%% EQ %%
\begin{equation}
	{C}_{\varphi,r}(x)=r_{20}x^2+r_{10}x+r_{00}
	\label{ec:Cr0-ejemploD}
\end{equation}
The equation \refp{ec:relacion-ejemploD} just establishes a relationship between $C_{\varphi,b}(x)$ and $C_{\varphi,r}(x)$.
Let's assume, arbitrarily, that $C_{\varphi,r}(x)$ is an already known polynomial, thus, $C_{\varphi,b}(x)$ becomes an unknown function.
In oder to simplify the notation, let's call $y(x)=C_{\varphi,b}(x)$, thus the characteristic equation \refp{ec:relacion-ejemploD} is particularised as:
%% EQ %%
\begin{equation}
	\frac{dy}{dx}-\frac{2}{x}y+\frac{2(r_{20}x^2+r_{10}x+r_{00})+(1-x)(2r_{20}x+r_{10})}{x}=0
\end{equation}
or more compactly:
%% EQ %%
\begin{equation}
	\begin{split}
	&\frac{dy}{dx}-\frac{2}{x}y+f(x)=0 \\
	&f(x)=(r_{10}+2r_{20})+\frac{2r_{00}+r_{10}}{x}
	\end{split}
	\label{ec:reladif-ejemploD}
\end{equation}
The general solution of this differential equation is (see the details in \refp{app:integracion}):
\begin{equation}
	y(x)={C}_{\varphi,b}(x)=b_{20}x^2+\underbrace{(r_{10}+2r_{20})}_{b_{10}}x+\underbrace{\frac{2r_{00}+r_{10}}{2}}_{b_{00}}
	\label{ec:Cb0-ejemploD}
\end{equation}
Looking at the two head \emph{loss} coefficient equations \refp{ec:Cr0-ejemploD} and \refp{ec:Cb0-ejemploD}, both contain six coefficients $\{ r_{20},r_{10},r_{00},b_{20},b_{10},b_{00}\}$ but only four of them are independent, $\{ r_{20},r_{10},r_{00},b_{20}\}$.
For this particular case $C_{junction}$, eq.\refp{ec:junction-x1}, has the form:
%% EQ %%
\begin{equation}
	{C}_{junction}=e_3x^3+e_2x^2+e_1x+e_0
	\label{ec:potdis-ejemploD}
\end{equation}
where the coefficients of eq.\refp{ec:potdis-ejemploD} are already known by measurements.
By using the expression ${C}_{junction}={C}_{\varphi,b}\cdot x+{C}_{\varphi,r}\cdot(1-x)$, along with the previous results, the following relationships hold:
%% EQ %%
\begin{equation}
	\begin{split}
		&e_3=b_{20}-r_{20}\\
		&e_2=3r_{20}\\
		&e_1=\frac{3}{2}r_{10}\\
		&e_0=r_{00}
	\end{split}
	\label{ec:coefdis-ejemploD}
\end{equation}
It is a system of four equations and four unknowns, thus clearing:
%% EQ %%
\begin{equation}
	\begin{split}
		&r_{20}=\frac{1}{3}e_2\\
		&r_{10}=\frac{2}{3}e_1\\
		&r_{00}=e_0\\
		&b_{20}=e_3+r_{20}
	\end{split}
	\label{ec:Cb0Cr0-ejemploD}
\end{equation}
This means that since $C_{junction}$ is an input datum, $C_{\varphi,r}$ is discovered with eqs.\refp{ec:Cb0Cr0-ejemploD} and by using eq.\refp{ec:Cb0-ejemploD} $C_{\varphi,b}$ can be obtained, as well.
In this way we would be able to discover how to split the energy dissipation at the junction into its \emph{internal} head \emph{loss} coefficients using the MinEDP and the measured $C_{junction}(x)$. 
This corresponds to the full dots in figure \refp{fig:exampleD-2}).

Now, it remains to find out the work exchanged, or equivalently, the work exchange coefficient $C_{\bar{d}}$.
There is still one measured input not used: the head \emph{change} coefficient in one branch.
Following the qualitative analysis of the previous section \S\ref{sec:Qualitative}, now we are going to compare the void dots with the full dots at the same $x$.
Let's assume we know the values at the side-branch and they are fitted to a $2^{nd}$ oder polynomial:
%% EQ %%
\begin{equation}
	{C}_{r}=r_2x^2+r_1x+r_0
	\label{ec:Cr-ejemploD}
\end{equation}
Therefore using eq.\refp{ec:C-ejemploD}:
%% EQ %%
\begin{equation}
	C_{\bar{d}}=\left({C}_{r,\varphi}-{C}_r\right)\cdot (1-x)=\left[ (r_{20}-r_2)x^2+(r_{10}-r_1)x+(r_{00}-r_0)\right]\cdot(1-x)
\end{equation}
As it can be seen, the `single' work exchange coefficient $C_{\bar{d}}$ is not a $2^{nd}$ order polynomial, but a $3^{rd}$ order one.
%% EQ %%
\begin{equation}
	C_{\bar{d}}=d_3x^3+d_2x^2+d_1x+d_0
\end{equation}
By identifying coefficients:
%% EQ %%
\begin{equation}
	\begin{split}
		&d_3=r_2-r_{20}\\
		&d_2=-(d_1+d_3)\\
		&d_1=r_{10}-r_1\\
		&d_0=0
	\end{split}
	\label{ec:D-ejemploD}
\end{equation}
Notice that at the boundaries of the flow distribution $x=0$ and $x=1$ the exchanged work is zero, i.e., $C_{\bar{d}}=0$.
Finally, with $C_{\varphi,b}$ eqs.\refp{ec:Cb0-ejemploD}, \refp{ec:C-ejemploD} and $C_{\bar{d}}$ the other \emph{external} head \emph{change} coefficient can be computed.
If measured then the value must match the computed one.
\newline

Summarizing, it has been proven that using as input the \emph{external} parameters \{${C}_{junction}$, ${C}_r$\} or \{${C}_{junction}$, ${C}_b$\} or equivalently the head \emph{change} coefficients \{${C}_b$,${C}_r$\} the three independent parameters needed for the \emph{internal} model of the junction \{ ${C}_{\varphi,r}$,${C}_{\varphi,b}$, $C_{\bar{d}}$\} can be computed.
The key is to add a new equation provided by the MinEDP.
In doing so, it has been found a \emph{characteristic differential equation} which relates the two head \emph{loss} coefficients.

One obstacle to check this method is the impossibility of measuring directly the \emph{internal} model.
In other words, although its works mathematically, some other evidence would be enlightening.
The CFD calculations based on fundamental laws may help.
Next section, applies the particular case developed here to a very concrete case.

%% SECTION %%%%%%%%%%%%%%%%%%%%%%%%%%%%%%%%%%%%%%%%%%%%%%%%%%%%%%%%%%%%%%%%%%%%%%%
\section{Estimation of $d$ by: CFD, experimentally and the MinEDP}
\label{sec:validation}
El aim of this section is to apply and to proof the previous method to obtain the parameters of the \emph{internal} model of a branched junction from \emph{external} measurements (i.e. head \emph{change} coefficients).
The comparison is made based on the available data in the literature; experimental data, an evaluation of $d$ performed by CFD and the previous method based on MinEDP.

Unfortunately, the experimentally $d$ cannot be measured (`directly'), but Herwig et al. \cite{SCHMANDT2015268} estimates the $d$ by using CFD.

\begin{figure}[h]
	\centering
	\includegraphics[width=0.85\textwidth]%
	{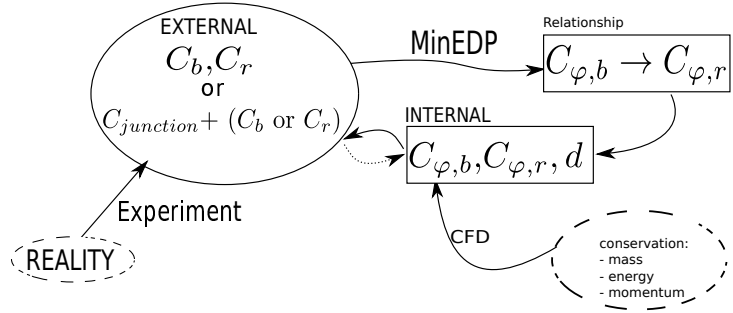}% picture filename
	\caption{Scheme of the comparison among experiment, CFD and the Principle of Minimum Dissipation.}
	\label{fig:CFD-ZHU-MinEP}
\end{figure}

Fortunately, our method can be applied to the same experimental case previously collected in Miller's book \cite{miller1990internal} and by Zhu's thesis \cite{ZhuThesis} which in turn, were also used by Herwig et al. \cite{SCHMANDT2015268} to check their CFD tests.
The validation strategy of Herwig et al., was to check if their computed head \emph{change} coefficients $C_b$ and $C_r$ agree with the measured experimental ones, obtained by Zhu and Miller.
If so, then Herwig et al. assumed that their computed $d$ values should be equal to the actual ones occurring at the Zhu's experimental tests.
Our strategy is to start from the experimental data of Zhu, and to compute the \emph{internal} parameters with the methodology developed previously.
Afterwards our \emph{internal} model is compared with Herwig's outcomes, obtained by using CFD.
In figure \refp{fig:CFD-ZHU-MinEP}, a scheme which summarises the strategies is shown.
%% FIG %%

\begin{figure}[h]
	\centering
	\includegraphics[width=1.0\textwidth]%
	{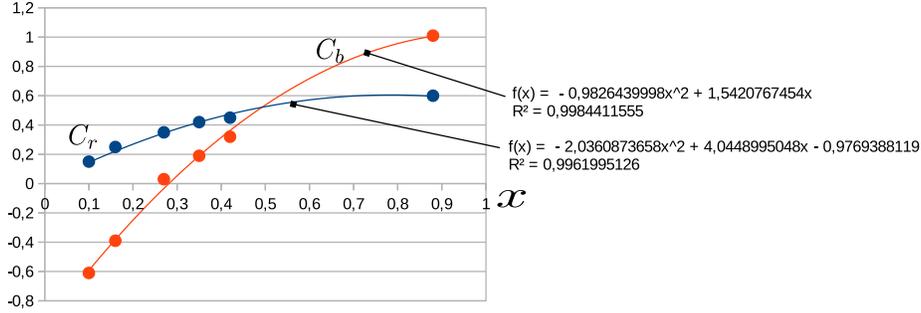}% picture filename
	\caption{(Circles) Zhu measured values. (Lines) polynomial fittings.}
	\label{fig:Zhu}
\end{figure}

\begin{figure}[!ht]
	\centering
	\includegraphics[width=0.85\textwidth]%
	{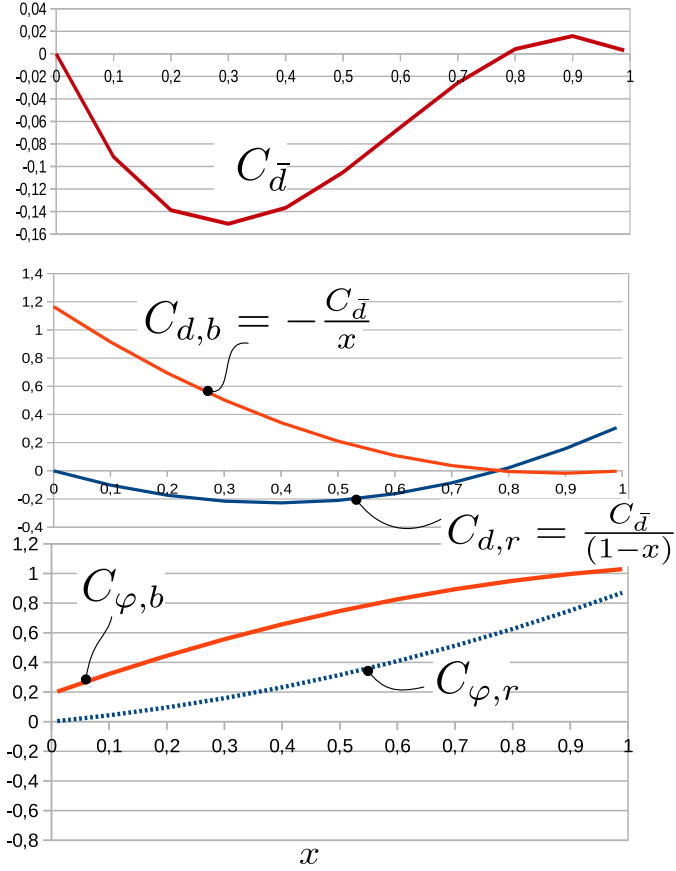}% picture filename
	\caption{(Top) Work exchanged per unit of kinetic energy at the common branch.(Middle) Diffusive work coefficients.(Bottom) Pure dissipative coefficients. Note: The difference between the bottom curve and the respective middle curve, gives the curves in figure \refp{fig:Zhu} for the side-branch $b$ and straight $r$ branches.}
	\label{fig:D-Zhu}
\end{figure}

The experimental case was a square cross-sectional convergent branched junction at $90^{\circ}$ (T-junction).
The cross-sectional areas of the straight, side-branch and common sections are the same.
Table \refp{tab:Zhu} displays the measured values by Zhu \cite{ZhuThesis}.
These values have been fitted to a $2^{nd}$ order polynomial and the result is shown in figure \refp{fig:Zhu}.
%% TAB %%
\begin{table}[ht]
	\centering
\begin{tabular}{lll}
	$x$ &	$C_r$ &	$C_b$	 	\\\hline
	0.10&0.15&-0.61\\
	0.16&0.25&-0.39\\
	0.27&0.35&0.03\\
	0.35&0.42&0.19\\
	0.42&0.45&0.32\\
	0.88&0.60&1.01
\end{tabular}
\caption{Zhu's thesis, measured values \cite{ZhuThesis} ref.(Page 141).}
\label{tab:Zhu}
\end{table}
%% FIG %%
Based on the polynomial fit of $C_b$ and $C_r$ and applying  $C_{junction}=C_b\cdot x+C_r\cdot(1-x)$, the following polynomial describes the energy dissipation coefficient of the T-junction:
%% EQ %%
\begin{equation}
	C_{junction}=\underbrace{-1.053443366}_{e_3}\cdot x^3+\underbrace{1.52017876}_{e_2}\cdot x^2+\underbrace{0.5651379331}_{e_1}x
	\label{ec:Cjunction-exampleD}
\end{equation}
Therefore the form of these equations allows the straightforward application of the particular case described in \S\ref{sec:particular}.
In concrete, our input (experimental) data are: equation \refp{ec:Cjunction-exampleD} and the $C_r(x)$ in figure \refp{fig:Zhu}.
First, let's use the coefficients from eq.\refp{ec:Cjunction-exampleD} in eqs. \refp{ec:Cb0-ejemploD} and \refp{ec:Cb0Cr0-ejemploD} to obtain the following set of coefficients:
%% EQ %%
\begin{equation}
	\begin{split}
	r_{20}&=\frac{1}{3}\cdot e_2=0.5067262532 \\
	r_{10}&=\frac{2}{3}\cdot e_1=0.3767586223 \\
	r_{00}&= e_0=0.0 \\
	b_{20}&= e_3+\frac{1}{3}\cdot e_2=-0.5467171128\\
	b_{10}&= 2\cdot r_{20}+r_{10}=1.3902111287\\
	b_{00}&= \frac{1}{2}\cdot r_{10}+r_{00}=0.1883793112
\end{split}
\end{equation}
Here the characteristic differential equation has been applied to find out  $C_{\varphi,b}(x)$ and $C_{\varphi,r}(x)$, the head \emph{loss} coefficients.
Recall that this split corresponds to the full dots of fig.\refp{fig:exampleD-2}.
Now, comparing the actual head \emph{change} coefficient with the head \emph{loss} coefficient (see eq.\refp{ec:D-ejemploD}), the `single' work exchange coefficient $C_{\bar{d}}$ is found:
%% EQ %%
\begin{equation}
	\begin{split}
		d_3&=r_2-r_{20}=-1.489370253\\
		d_2&=-(d_3+d_1)=2.6546883761\\
		d_1&= r_{10}-r_1=-1.1653181231\\
		d_0&=0.0
	\end{split}
	\label{ec:D-Zhu}
\end{equation}
Figure \refp{fig:D-Zhu} shows our output parameters: the head \emph{loss} coefficients $C_{\varphi,b}$, $C_{\varphi,r}$, the `single' work exchange coefficient $C_{\bar{d}}$ and, in turn, this latter can be split into the work exchange coefficients at each branch $C_{d,b}$ and $C_{d,r}$.
%% FIG %%
Notice that $C_{\varphi,b}$, $C_{\varphi,r}$ are always positive, as was to be expected for a head \emph{loss} coefficient but also they are completely different from the head \emph{change} coefficients measured and displayed in fig. \refp{fig:Zhu}
The `single' work coefficient is zero at $x=0$ and $x=1$ as was expected since there there is no work interaction between the streams.
The physical interpretation of the $C_{\bar{d}}$ curve is: from $x=0$ to around $x=0.8$, due to its negative sign, work is transferred from the strait branch towards the side-branch and from that $x=0.8$ until $x=1$ the side-branch transfers diffusively work towards the strait branch.
Intuitively, in the first part, the flow rate at the strait branch is much higher than in the side branch while in the second part is just the opposite, but its interaction is much less.
By adding both effects, the dissipation and the work transfer, the \emph{external} model is obtained.
In concrete using the relationships eq.\refp{ec:varphi-daga} and \refp{ec:varphi-dagger-1}, the head \emph{change} coefficients measured by Zhu are recovered.
Approximately, in the range $x\in (0,0.3)$ the $C_b$ becomes negative.
The physical interpretation is that the work transfer from the strait branch $r$ to the side-branch $b$ is bigger than the energy dissipation at that latter branch $b$, and this phenomenon shows up, traditionally, as a `negative loss'.
However, notice that within the range $x\in (0.3,0.8)$, both head \emph{change} coefficients (those measured by Zhu), are positive and it is easy to think, mistakenly,  that there is no work interaction, but as fig. \refp{fig:D-Zhu} shows, it is not true.
The work is only zero at around $x=0.8$ and only, at this flow conditions,  $C_b=C_{\varphi,b}$ and $C_r=C_{\varphi,r}$.

Our last step is to compare our results with those obtained by CFD by Herwig et al. \cite{SCHMANDT2015268}, for the same case as Zhu.
Before going into the details, a caveat is needed.
By construction, our results shown in fig.\refp{fig:D-Zhu} reproduce exactly (to the approximated fit) the measured values of Zhu.
However, the information from Herwig et al. has additional uncertainties, as they also mention.
For instance, there is lack of data about the accurate geometry of the Zhu's T-junction.
Therefore, Herwig's CFD model may have additional uncertainties with respect to Zhu's steup.

Herwig presented the results as the ratios ${d_b}/{\varphi_b}$ and ${d_r}/{\varphi_r}$, which are equivalent to the ratios of the \emph{internal} coefficients $(C_{d,b}/C_{\varphi,b})$ and $(C_{d,r}/C_{\varphi,r})$, respectively. 
%% FIG %%
\begin{figure}[!ht]
	\centering
	\includegraphics[width=0.85\textwidth]%
	{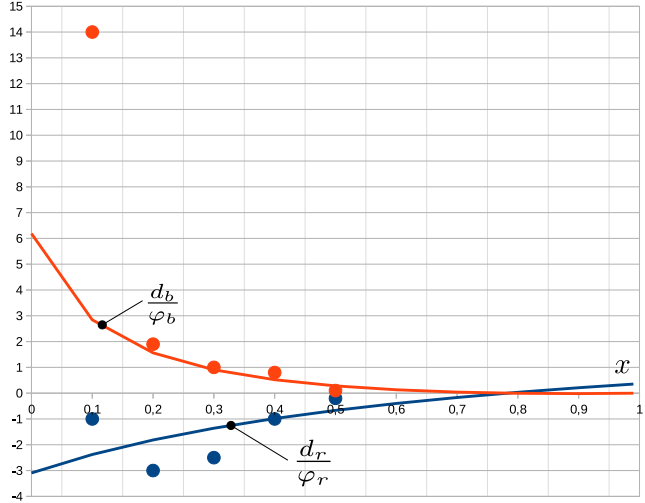}% picture filename
	\caption{Comparison between the ratios $(d_b/\varphi_b)$ and $(d_r/\varphi_r)$ computed by Herwig et al. \cite{SCHMANDT2015268} (bullets) and estimated using the Minimum Dissipation Principle and an analytical approach (lines).}
	\label{fig:Herwig}
\end{figure}
Figure \refp{fig:Herwig} shows the comparison between MinEDP and the CFD.
The five bullets represent the Herwig's CFD computations, while the lines are the results from MinEDP.
In our opinion, the agreement is quite good or even surprisingly good, taking into account the big difference between the CFD and MinEDP methods employed.
The CFD data reported by Herwig was computed on a grid with $13,036,950$ cells.
According to fig.$(4)$ in \cite{SCHMANDT2015268}, for $x<0.4$ the agreement of the computed $C_b$ and $C_r$ by CFD and the experimental values of the literature is no so good.
Herwig tried solve the disagreement in their fig.$(5)$ \cite{SCHMANDT2015268}, by adjusting the kinetic energy correction factor $\alpha$ in the definition of the computed $C_b$ and $C_r$.
Obviously, this adds uncertainty to the comparison.
In fact, for $x<0.4$, fig. \refp{fig:Herwig} shows an increasing disagreement between the CFD and MinEDP.
The CFD point most to the left in fig. \refp{fig:Herwig} is completely in disagreement.
Fortunately, Herwig et al. provided much more information about their CFD computations.
In their appendix A: Numerical details \cite{SCHMANDT2015268}, they state:
\begin{quote}
{For grid number 2 the $K_{23}$-value (our $C_b$), however, shows an irregular behavior for $r_{\dot{m}}=0.1$ (our $x=0.1$)}
\end{quote}
, which is precisely where a strong disagreement appears.
It seems that, for low $x$ values, i.e., low flow rate through the side-branch, the CFD calculations became unstable in that branch.
Finally the CFD was performed within the range $x\in (0.1,0.5)$ while the MinEDP provides information over the whole range, thus making clear the evolution of the work exchange with $x$.
\linebreak

To illustrate how powerful, practical and enlightening this new method can be, in \ref{annex:example} the previous particular case, exposed in section \ref{sec:particular}, has been applied to a recent experimental work \cite{ZIGANSHIN2021107491}.
Take into account that the annex deals with ten cases (five configurations times two cases) analysed for the whole $x$ range.
The cost of CFD analysis would be impractical.

\section{Conclusions}
This paper makes use of a new method developed by the authors in three previous works \cite{SOTOFRANCES2019181} \cite{HEFAT2019} \cite{SOTO2021-MinEDP} to find out the steady-state of a flow network.
The new method is based on the Minimum Energy Dissipation Principle (MinEDP).

Herwig and Schmandt \cite{SCHMANDT2015268} showed, by usign CFD, that the \emph{conventional head loss coefficients} may become negative due to an internal diffusive work exchange which they named $d$.
These \emph{negative loss} has originated much controversy for long.
Stigler \cite{jaroslavstigler:2006:1} was very critical and proposed to use physically meaningful parameters.
In this respect, the new perspective provided by Herwig and Schmandt leads to what here has been called the \emph{internal} model of a branched junction, in opposition to the conventional model which here is called \emph{external} or measurable model.
The idea of Herwig and Schmandt was to separate \emph{pure} energy dissipation from the possible work exchange within a branched junction.
Thus, fulfilling the Stigler's desire for a physically meaningful definition.
However, this well behaved split, needs three \emph{internal}, i.e., not directly measurable parameters, while the \emph{external} model needs only two.
Therefore for very long the conventional \emph{head loss coefficients} have been the only choice, in engineering practice.
However, as it was shown in \cite{SOTO2021-MinEDP}, overlooking the paradoxical definition of a possible \emph{negative loss} generates much confusion and misunderstanding on the engineering practitioners.

Taking advantage of the new insights provided by the application of the MinEDP, this paper answers affirmatively to the following question: is it possible to get the \emph{internal} model of a branched junction from \emph{external} measurements?.
The MinEDP showed that the steady-state of a flow network has different properties depending on the existence of work exchange at the junctions or not.
In the first case the steady-state is not a minimum of the global dissipation function, while in the second it does coincide with the minimum.
Since the \emph{external} model has two parameters while the \emph{internal} has three, the previous property of the steady-state provided a hint for the solution: the missing equation could be found in the MinEDP.
This paper corroborates that.
Moreover, an unexpected and outstanding outcome of the paper is the discovery of the general characteristic differential equation for any branched junction.
This equation relates the \emph{pure} energy dissipation at both branches, i.e., establishes a relationship between $C_{\varphi,r}$ and $C_{\varphi,b}$.

The paper compares the Zhu's experimental results for a convergent T-junction, with Herwig's CFD calculations and the MinEDP.
By construction MinEDP completely agrees with the experimental values from an \emph{external} standpoint.
Thus, the possibility of confirming our results by using CFD, has also been carried out based on \cite{SCHMANDT2015268}.
The final conclusion is that, taking into account the uncertainties and the huge differences between the CFD and MinEDP methods, the agreement is surprisingly good.

The MinEDP method provides an economic and reliable way to compute $d$ for the whole range of the side-branch flow ratios.
One practical application of the method could be improving the performance of central air exhaust ventilation systems \cite{TONG2019134}.
The branched junctions could be designed to minimise their work interaction thus leading to duct networks less sensible to on-site changes of the hydraulic resistances (see \cite{SOTO2021-MinEDP}) or changes in their operation.

%% APENDICE %%%%%%%%%%%%%%%%%%%%%%%%%%%%%%%%%%%%%%%%%%%%%%%%%%%%%%%%%%%%%%%%%%%%%%
\appendix
\section{Integration of the characteristic differential equation. Particular case.}
\label{app:integracion}
The solution of the linear non-homogeneous differential equation $(y'=dy/dx)$:
 \begin{equation}
 y'+P(x)y=Q(x)
 \end{equation}
can be written as:
 \begin{equation}
 y=y_0\cdot C+y_p
 \label{ec:y}
 \end{equation}
where $y_0$ is the solution to the homogeneous equation: $y'+P(x)y=0$ and $C$ is a constant.

In order to find a particular solution, an integrating factor must be sought:
 $$
 \frac{d}{dx}(f\cdot g)=f\cdot g'+f'\cdot g
 $$
Therefore there must exist a function $\mu$ such that:
 $$
 \mu\cdot y'+\mu' \cdot y=\frac{d}{dx}(\mu\cdot y)
 $$
since:
 $$
 \frac{d}{dx}(\mu\cdot y)=\mu\cdot y'+\mu\cdot P(x) \cdot y= \mu\cdot Q(x)
 $$
therefore:
 $$
 y_p=\frac{1}{\mu}\cdot \left( \int{\mu \cdot Q(x) dx}+C_i\right)
 $$
a $\mu$ is needed that fulfills:
 $$
 \mu\cdot P(x)=\mu'
 $$
thus:
 $$
 \int{\frac{\mu'}{\mu}\cdot dx}=\int{P(x)\cdot dx}
 $$
integrating:
 $$
 \mu(x)=e^{\int{P(x)\cdot dx}}
 $$
For the particular case of the paper's equation:
 $$
 y'-\frac{2}{x}\cdot y=-(r_{10}+2r_{20})-\frac{2r_{00}+r_{10}}{x}=Q(x)
 $$
 $$ P(x)=-\frac{2}{x} $$
The solution to the homogeneous part is:
 $$
 y_0'-\frac{2}{x}\cdot y_0=0
 $$
 $$
 y_0(x)=C_0\cdot x^2
 $$
The integrating factor of the particular solution is:
 $$
\mu=e^{-\int{\frac{2}{x}dx}}=e^{-2\ln|x|}=\frac{1}{x^2}
$$
Therefore $y_p$ has the form:
%% EQ %%
\begin{eqnarray}
y_p =& x^2\cdot\left( \int{-\frac{r_{10}+2r_{20}}{x^2}-\frac{(2r_{00}+r_{20})}{x^3}\cdot dx}+C_i \right)\\
y_p =& x^2\cdot\left( +\frac{(r_{10}+2r_{20})}{x}+\frac{1}{2}\frac{(2r_{00}+r_{20})}{x^2}+C_i \right)\\
y_p =& C_ix^2+(r_{10}+2r_{20})x+\frac{1}{2}(2r_{00}+r_{20})
\end{eqnarray}
by using eq.\refp{ec:y} and renaming the coefficients as:
%% EQ %%
\begin{eqnarray}
C_0\cdot C+C_i=& b_{20}\\
(r_{10}+2r_{20})=& b_{10}\\
\frac{1}{2}(2r_{00}+r_{20})=& b_{00}
\end{eqnarray}
the $y(x)$ solution equation \refp{ec:Cb0-ejemploD} is obtained.

\section{Example of application to a novel exhaust duct device}
\label{annex:example}
\begin{figure}[!ht]
	\centering
	\includegraphics[width=0.80\textwidth]%
	{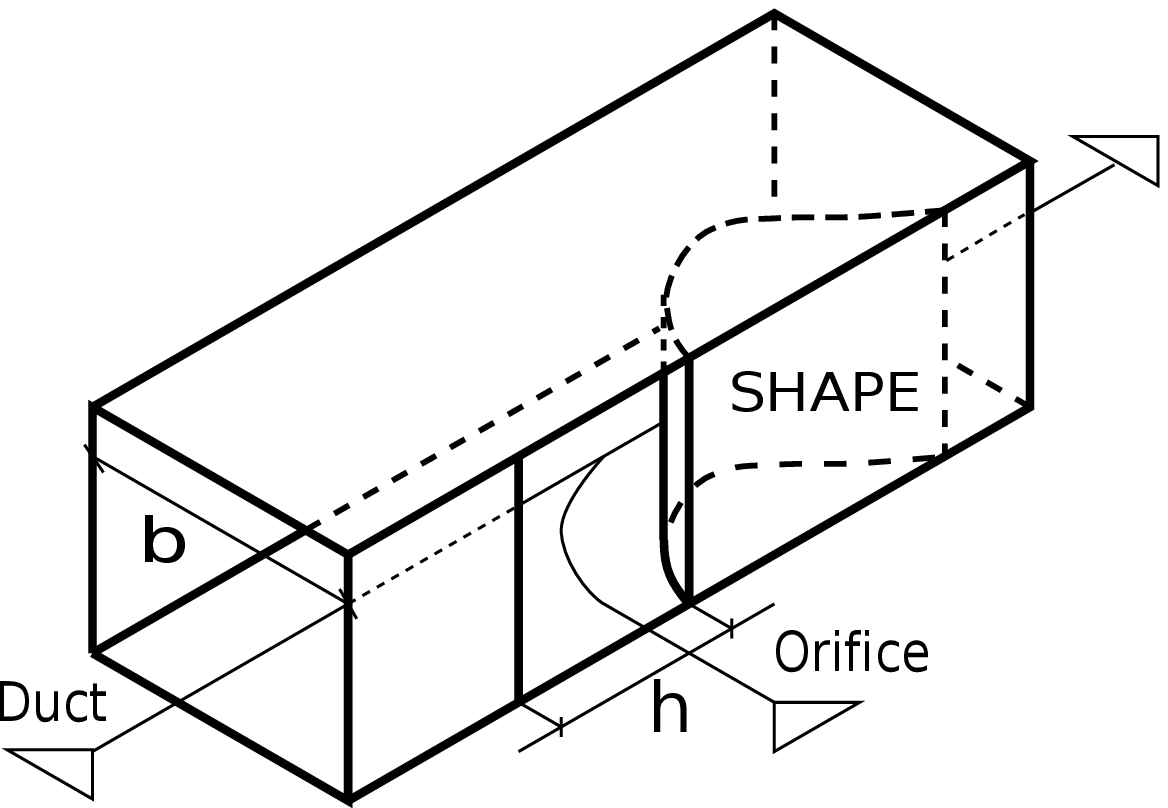}% picture filename
	\caption{Scheme of midpoint lateral orifice of the exhaust air duct studied by Ziganshin et al. }
	\label{fig:shaped}
\end{figure}

Recently Ziganshin et al. \citep{ZIGANSHIN2021107491} studied how to reduce what they called:
``\emph{the drag of midpoint lateral orifices of exhaust air ducts by shaping them along vortex zone outlines.}''
Usually in fluid mechanics, by \emph{drag}, it is understood the friction or resistance to flow, but this is misleading in branched junctions as \cite{SOTO2021-MinEDP} shows.
Ziganshin named $C_r$ (duct) and $C_b$ (orifice) as LDC the local drag coefficients and named them $\zeta_P$ and $\zeta_O$ respectively.
The widespread name is \emph{the head loss coefficient}.
Figure \refp{fig:shaped} shows the type of exhaust duct studied.
There are two duct types: without the profile (\emph{no-shape}) and the profiled (shaped) one.
The goal of the shape is to reduce the dissipation mechanisms due to the vortex at the entrance of the orifice (see \citep{ZIGANSHIN2021107491} for details).
%% FIG %%
Ziganshin fitted both, $C_r$ and $C_b$, to a quadratic polynomial for the case of no-shape ($C_{r,nsh}$ $C_{b,nsh}$) and shape ($C_{r,sh}$ $C_{b,sh}$).
Table \ref{tab:n-shape} reproduces Table 3 of \citep{ZIGANSHIN2021107491} for the no-shape case.
Table \ref{tab:shape} reproduces Table 4 of \citep{ZIGANSHIN2021107491} for the shape case.
%% TABLE %%
\begin{table}[]
\centering
\begin{tabular}{llllllll}
\hline
\multicolumn{4}{c}{$C_{r,nsh}=A\cdot x^2+B\cdot x+ C$ } & \multicolumn{4}{c}{$C_{b,nsh}=A\cdot x^2+B\cdot x+ C$ }  \\ \cline{1-4}\cline{6-8}
 $h/b$   &  A   & B    & C   &     &  A   & B  & C \\ \cline{1-4}\cline{6-8}
 $0.32$   & -0.9128 & 2.1585 & -0.0481    &     & 21.592 & 0.9244 & -0.6842 \\
 $0.60$   & -0.9761	& 2.1186 & -0.0447    &     & 4.6088 & 2.1499 &	-0.7474 \\
 $1.00$   & -1.0316	& 2.1407 & -0.0453    &     & 0.1298 & 3.2531 &	-0.9285 \\
 $1.50$   & -1.0667	& 2.1713 & -0.0531    &     & -1.3010 &	3.8009 & -1.0118 \\
 $2.00$   & -1.0740 & 2.1721 & -0.0525    &     & -1.6854 &	3.9822 & -1.0497 \\\hline 
\end{tabular}
\caption{Ziganshin fit of the measured head \emph{loss} coefficients. NO-SHAPE}
\label{tab:n-shape}
\end{table}
%% TABLE/ %%
%% TABLE %%
\begin{table}[]
\centering
\begin{tabular}{llllllll}
\hline
\multicolumn{4}{c}{$C_{r,sh}=A\cdot x^2+B\cdot x+ C$ } & \multicolumn{4}{c}{$C_{b,sh}=A\cdot x^2+B\cdot x+ C$ }  \\ \cline{1-4}\cline{6-8}
 $h/b$   &  A   & B    & C   &     &  A   & B  & C \\ \cline{1-4}\cline{6-8}
 $0.32$   & -7.7828 & 2.9432 & -0.0041    &     & 11.654 & 4.7656 & -0.869 \\
 $0.60$   & -1.7804	& 1.6072 &  0.0412    &     & 2.9347 & 2.1304 &	-0.7175 \\
 $1.00$   & -1.0661	& 1.6751 &  0.0197    &     & 0.2212 & 2.5399 &	-0.7175 \\
 $1.50$   & -0.8107	& 1.5291 &  0.0066    &     & -1.1494 &	3.2512 & -0.9666 \\
 $2.00$   & -0.8721 & 1.6813 &  0.0022    &     & -1.5356 &	3.5147 & -0.9881 \\\hline 
\end{tabular}
\caption{Ziganshin fit of the measured head \emph{loss} coefficients. SHAPE}
\label{tab:shape}
\end{table}
%% TABLE/ %%
%% FIG %%
\begin{figure}[!ht]
	\centering
	\includegraphics[width=0.80\textwidth]%
	{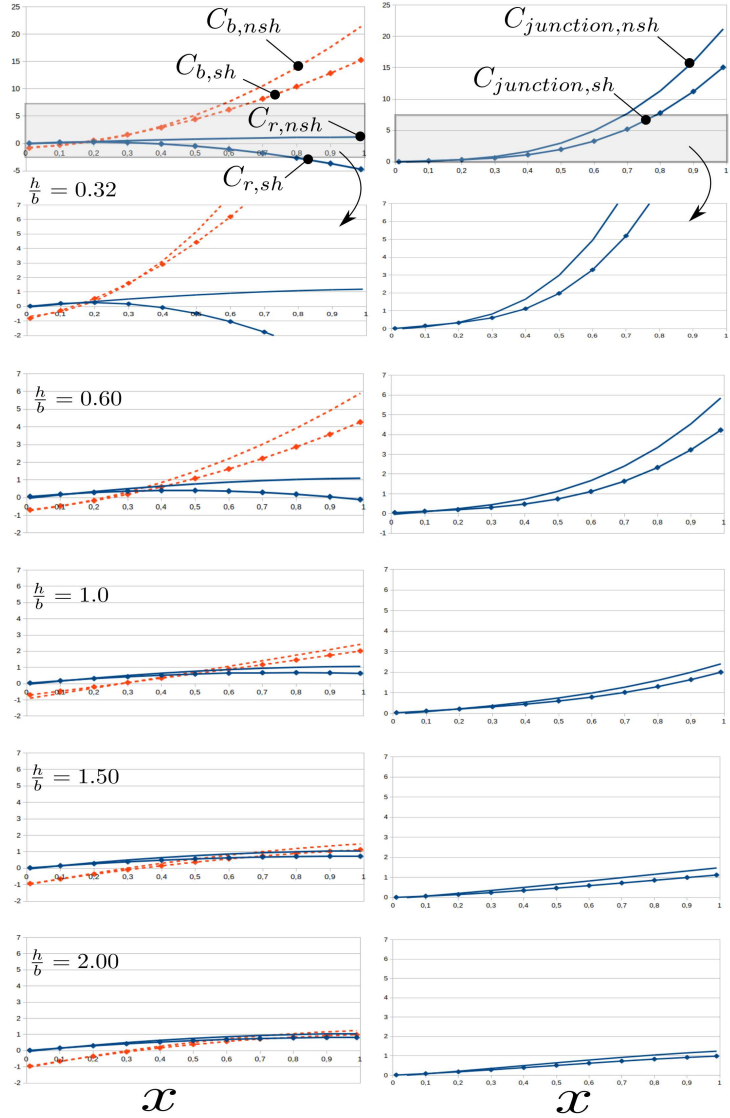}% picture filename
	\caption{Ziganshin et al. measurements. Left: $C_r$ and $C_b$. Right: $C_{junction}$ or dissipation coefficient of the junction. \emph{Note: no-shape case (lines), shape case (lines-dots)} }
	\label{fig:CrCbDissp}
\end{figure}
On the left side of figure \refp{fig:CrCbDissp} the curves of tables \ref{tab:n-shape} and \ref{tab:shape} are represented.
On the right, the dissipation coefficient of the junction (using eq.\refp{ec:Cjunction}) is shown.
Clearly $C_{junction}$ must be always positive and figure \refp{fig:CrCbDissp} also makes clear that for the same flow distribution $x$, the shape case dissipates less energy.

\begin{figure}[!ht]
	\centering
	\includegraphics[width=0.80\textwidth]%
	{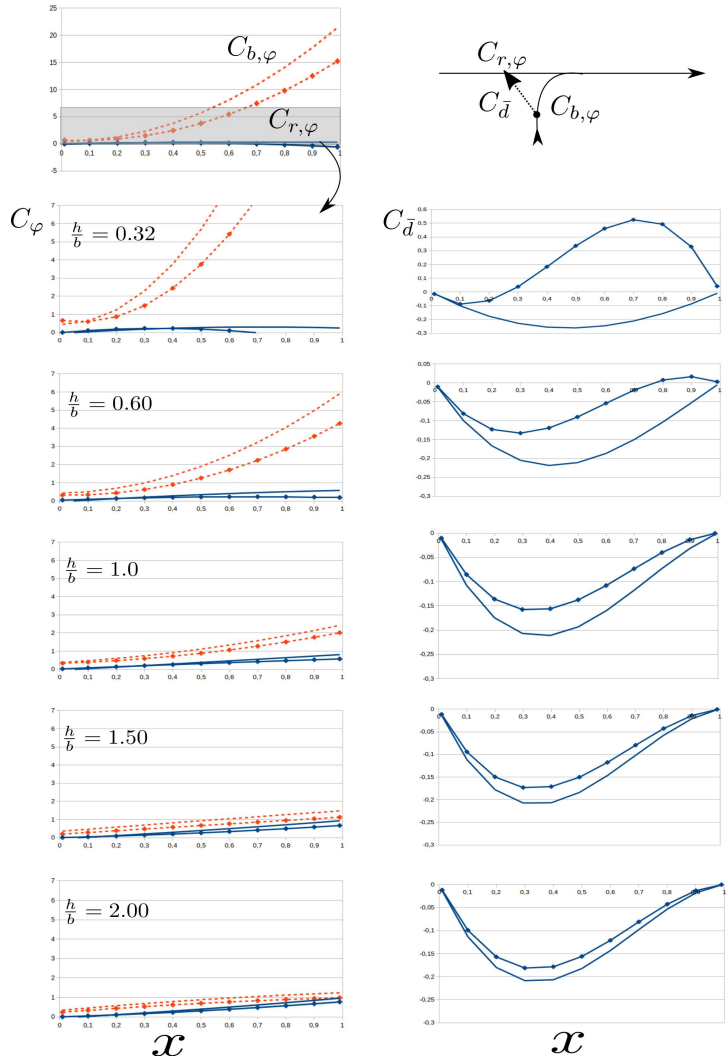}% picture filename
	\caption{Internal model obtained using the MinEDP. Left: pure head loss coefficients Right: diffusive shear work coefficient. \emph{Note: no-shape case (lines), shape case (lines-dots)} }
	\label{fig:CphiCd}
\end{figure}

Using the results of section \ref{sec:particular} the three parameters of the internal model for both no-shape and shape cases, can be discovered, i.e.: the \emph{pure} head loss coefficients $C_{r,\varphi}$ and $C_{b,\varphi}$ (tables \ref{tab:n-shape-internal} and \ref{tab:shape-internal} ) and the work coefficient $C_{\bar{d}}$ (table \ref{tab:d}).
Figure \refp{fig:CphiCd} shows graphically the equations for the internal model.
They provide much more information than the conventional external model (figure \refp{fig:CrCbDissp}).
For instance, in the extrema case $h/b=0.32$, $C_{r,\varphi,sh}$ becomes negative for $x>0.7$ and this flags to correct slightly the fitting for that case.
Another interesting point is that the scale of the shear work coefficient does not vary as much as the dissipative coefficients.
When there is no shape, $C_{\bar{d}}$ is always negative, that means that work is transferred from the duct to the stream coming from the orifice (side-branch).
However, this is not true for the shape case.
Cases $h/b=0.32$ and $h/b=0.6$ indicate a transfer from the side-branch to the duct stream at high $x$ ratios.
There is an inversion in the work sense.
The effect is very strong for the extreme case $h/b=0.32$.
For the rest of cases, the work interaction is reduced in the shape case.
It is noteworthy the great change in form of the $C_{\bar{d}}$ coefficients.
As it was proven in \cite{SOTO2021-MinEDP}, the shear work interactions act as constraints, therefore if a duct network, with small $h/b$ ratios, was modified by placing these shape elements then, very likely, it would have a quite different, than the original, flow distribution.

\begin{figure}[!h]
	\centering
	\includegraphics[width=0.80\textwidth]%
	{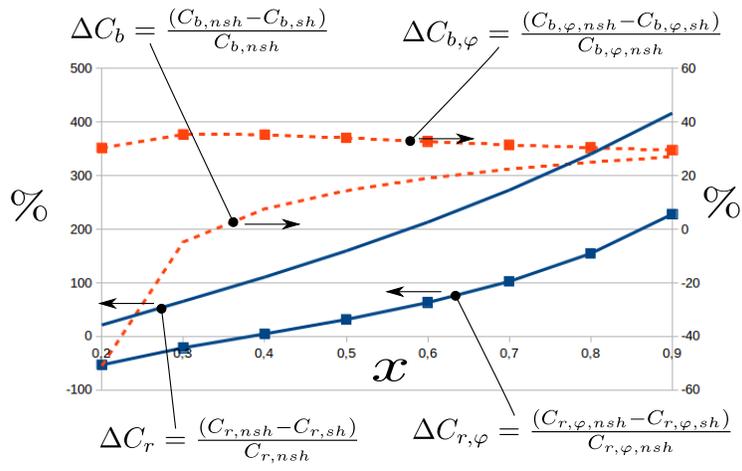}% picture filename
	\caption{Case $h/b=0.32$. Comparison of the reduction of the drag coefficient , as Ziganshin's Figure 16 \cite{ZIGANSHIN2021107491} (lines),  and the \emph{pure} head loss coefficient (lines-squares).}
	\label{fig:comparacion}
\end{figure}

Finally, this hidden shear work interaction currently leads to confusion since, mistakenly, one may think that the dissipation in one branch is reduced when there is, in fact, a work transfer from the other and vice versa.
Ziganshin et al. in Figure 16 of \cite{ZIGANSHIN2021107491} compared $\zeta_P$ and $\zeta_O$ (our $C_r$ and $C_b$) coefficients in both cases by estimating the percentage $\Delta \zeta$ of reduction of the drag.
Unfortunately, due to the aforementioned reason, this creates a misleading comparison, if the aim is to check the energy dissipation reduction.
Figure \ref{fig:comparacion} reproduces the Ziganshin's Figure 16 and compares the result with the reduction of the \emph{pure} (real) head loss coefficient.
In other words $\Delta C_r$ and $\Delta C_b$ are equal to $\Delta \zeta_P$ and $\Delta \zeta_O$ in \cite{ZIGANSHIN2021107491} but the real reduction in the dissipation mechanisms,  $\Delta C_{r,\varphi}$ and $\Delta C_{b,\varphi}$, are also shown.
The differences are noteworthy.
The reduction in the physical dissipation mechanisms in the flow through the side-branch (orifice) are approximately constant (around $30\%$) and independent of $x$, while the reduction of the drag goes to $0$ at $x=0.2$.
The physical interpretation would be that the shape reduces the dissipation induced by the vortex ``regardsless'', to some extent, of the $x$ value.
On the straight-branch (duct), Ziganshin's drag comparison (by implicitly assuming that drag equals resistance) seems to indicate a reduction of the dissipation for any $x$ reaching a peak reduction of $420\%$ at $x=0.9$.
However, it is not so.
In the range $x\in [0,0.35]$ there is no reduction in the dissipation mechanism, on the contrary, there is an increase (around $50\%$).
This seems logical, since the duct flow has the shape obstacle which was not present originally.
For $x>0.35$ the energy dissipation is effectively, reduced but the reduction is only around $220\%$, not $420\%$, almost one half.
%% TABLE %%
\begin{table}[]
\centering
\begin{tabular}{llllllll}
\hline
\multicolumn{4}{c}{$C_{r,\varphi,nshape}=r_{20}\cdot x^2+r_{10}\cdot x+ r_{00}$ } & \multicolumn{4}{c}{$C_{b,\varphi,nshape}=b_{20}\cdot x^2+b_{10}\cdot x+ b_{00}$ }  \\ \cline{1-4}\cline{6-8}
 $h/b$   &  $r_{20}$   & $r_{10}$    & $r_{00}$   &     &  $b_{20}$   & $b_{10}$  & $b_{00}$ \\ \cline{1-4}\cline{6-8}
 $0.32$   & -0.7156 & 1.0149 & -0.0481    &     & 21.7892 & -0.4163 & 0,4594 \\
 $0.60$   & -0.3149	& 0.9439 & -0.0447    &     &  5.2700 &  0.3141 & 0.4273 \\
 $1.00$   &  0.0269	& 0.8383 & -0.0453    &     &  1.1883 &  0.8922 & 0.3739 \\
 $1.50$   &  0.1876	& 0.8084 & -0.0531    &     & -0.0467 &	 1.1837 & 0.3511 \\
 $2.00$   &  0.2454 & 0.7833 & -0.0525    &     & -0.3660 &	 1.2740 & 0.3391 \\\hline 
\end{tabular}
\caption{Internal model: \emph{Pure} head loss coefficients based on MinEDP. NO-SHAPE}
\label{tab:n-shape-internal}
\end{table}
%% TABLE/ %%

%% TABLE %%
\begin{table}[]
\centering
\begin{tabular}{llllllll}
\hline
\multicolumn{4}{c}{$C_{r,\varphi,shape}=r_{20}\cdot x^2+r_{10}\cdot x+ r_{00}$ } & \multicolumn{4}{c}{$C_{b,\varphi,shape}=b_{20}\cdot x^2+b_{10}\cdot x+ b_{00}$ }  \\ \cline{1-4}\cline{6-8}
 $h/b$   &  $r_{20}$   & $r_{10}$    & $r_{00}$   &     &  $b_{20}$   & $b_{10}$  & $b_{00}$ \\ \cline{1-4}\cline{6-8}
 $0.32$   & -1.9868 & 1.3855 & -0.0041    &     & 17.4500 & -2.5881 & 0.6887 \\
 $0.60$   & -0.4191	& 0.5657 &  0.0412    &     &  4.2960 & -0.2725 & 0.3240 \\
 $1.00$   & -0.0671 & 0.6253 &  0.0197    &     &  1.2202 &  0.4911 & 0.3323 \\
 $1.50$   &  0.3038	& 0.3706 &  0.0066    &     & -0.0349 &	 0.9782 & 0.1919 \\
 $2.00$   &  0.3204 & 0.4607 &  0.0022    &     & -0.3431 &	 1.1015 & 0.2325 \\\hline 
\end{tabular}
\caption{Internal model: \emph{Pure} head loss coefficients based on MinEDP. SHAPE}
\label{tab:shape-internal}
\end{table}
%% TABLE/ %%

%% TABLE Cd%%
\begin{table}[]
\centering
\begin{tabular}{llllllll}
\hline
\multicolumn{4}{c}{$C_{\bar{d},nshape}=d_3\cdot d^3+d_{2}\cdot x^2+d_{1}\cdot x+ d_{0}$ } &  \multicolumn{4}{c}{$C_{\bar{d},shape}=d_3\cdot d^3+d_{2}\cdot x^2+d_{1}\cdot x+ d_{0}$ } \\
$h/b$   &  $d_{3}$   & $d_2$    & $d_1$ & & $d_{3}$   & $d_2$    & $d_1$\\ \hline
$0.32$ & -0.1972 & 1.3407 & -1.1436 & & -5.7960 & 7.3537 & -1.5577\\
$0.60$ & -0.6612 & 1.8358 & -1.1747 & & -1.3613 & 2.4029 & -1.0415\\
$1.00$ & -1.0585 & 2.3609 & -1.3024 & & -0.9990 & 2.0488 & -1.0498\\
$1.50$ & -1.2543 & 2.6172 & -1.3629 & & -1.1145 & 2.2730 & -1.1585\\
$2.00$ & -1.3194 & 2.7082 & -1.3888 & & -1.1925 & 2.4132 & -1.2206\\
 \hline
\end{tabular}
\caption{Internal model: Shear work exchange coefficient $C_{\bar{d}}$ based on MinEDP. NO-SHAPE and SHAPE. \emph{Note: $d_0=0$}}
\label{tab:d}
\end{table}
%% TABLE/ %%
%% FIG %%
%%
%% FIG %%

\afterpage{\clearpage}
\pagebreak
%%%%%%%%%%%%%%%%%%%%%%%%%%%%%%%%%%%%%%%%%%%%%%%%%%%%%%%%%%%%%%%%%%%%%%%%%%%%%%%%%%%%%%%%%%%%%%%%%%%
\section*{Acknowledgements}
\section*{References}
\bibliography{mybibfile_art}

\begin{thebibliography}{10}
\expandafter\ifx\csname url\endcsname\relax
  \def\url#1{\texttt{#1}}\fi
\expandafter\ifx\csname urlprefix\endcsname\relax\def\urlprefix{URL }\fi
\expandafter\ifx\csname href\endcsname\relax
  \def\href#1#2{#2} \def\path#1{#1}\fi

\bibitem{wood:1993}
J.~E.~F. Don J.~Wood, L. Srinivasa~Reddy, Modeling pipe networks dominated by
  junctions, Journal of Hydraulic Engineering 119~(8) (1993) 949--958.

\bibitem{jamesliggett:1994}
J.~A. Liggett, Discussion: Modeling pipe networks dominated by junctions,
  Journal of Hydraulic Engineering 120~(12) (1993) 1486--1489.

\bibitem{wood:1994}
J.~E.~F. Don J.~Wood, L. Srinivasa~Reddy, Clousure on: Modeling pipe networks
  dominated by junctions, Journal of Hydraulic Engineering 120~(12) (1994)
  1492--1493.

\bibitem{jaroslavstigler:2006:1}
J.~\u{S}tigler, Tee junction as a pipeline net element. part 1 a new
  mathematical model., Strojnícky časopis 57~(5) (2006) 249--262, iSSN: 0039-
  2472.

\bibitem{SCHMANDT2014191}
B.~Schmandt, V.~Iyer, H.~Herwig,
  \href{https://www.sciencedirect.com/science/article/pii/S0009250914000992}{Determination
  of head change coefficients for dividing and combining junctions: A method
  based on the second law of thermodynamics}, Chemical Engineering Science 111
  (2014) 191--202.
\newblock \href {http://dx.doi.org/https://doi.org/10.1016/j.ces.2014.02.035}
  {\path{doi:https://doi.org/10.1016/j.ces.2014.02.035}}.
\newline\urlprefix\url{https://www.sciencedirect.com/science/article/pii/S0009250914000992}

\bibitem{SCHMANDT2015268}
B.~Schmandt, H.~Herwig,
  \href{http://www.sciencedirect.com/science/article/pii/S0142727X15000697}{The
  head change coefficient for branched flows: Why “losses” due to junctions
  can be negative}, International Journal of Heat and Fluid Flow 54 (2015) 268
  -- 275.
\newblock \href
  {http://dx.doi.org/https://doi.org/10.1016/j.ijheatfluidflow.2015.06.004}
  {\path{doi:https://doi.org/10.1016/j.ijheatfluidflow.2015.06.004}}.
\newline\urlprefix\url{http://www.sciencedirect.com/science/article/pii/S0142727X15000697}

\bibitem{SOTOFRANCES2019181}
V.-M. Soto-Frances, J.-M. Pinazo-Ojer, E.-J. Sarabia-Escriva, P.-J.
  Martinez-Beltran, On using the minimum energy dissipation to estimate the
  steady-state of a flow network and discussion about the resulting
  power-law:application to tree-shaped networks in hvac systems, Energy 172
  (2019) 181--195,
  \url{https://www.sciencedirect.com/science/article/pii/S0360544219300623}.
\newblock \href
  {http://dx.doi.org/https://doi.org/10.1016/j.energy.2019.01.060}
  {\path{doi:https://doi.org/10.1016/j.energy.2019.01.060}}.

\bibitem{HEFAT2019}
V.-M. Soto-Frances, J.~manuel Pinazo-Ojer, E.-J. Sarabia-Escriva, P.-J.
  Martinez-Beltran, About using the minimum energy dissipation to find the
  steady-state flow distribution in networks,
  \url{https://advanceseng.com/new-pipe-network-analysis-method-minimum-energy-dissipation-principle-application-tree-shaped-duct-networks-hvac-systems/},
  presented at Heat Transfer, Fluid Mechanics and Thermodynamics, 14th
  international conference. HEFAT, Wicklow Ireland. (2019).

\bibitem{SOTO2021-MinEDP}
V.-M. Soto-Francés, J.-M. Pinazo-Ojer, E.-J. Sarabia-Escrivá,
  J.~Navarro-Esbrí,
  \href{https://www.sciencedirect.com/science/article/pii/S037877882100788X}{A
  new hvac ductwork steady-state flow analysis method: The minimum energy
  dissipation principle applied to flow networks including the effects of
  branched junctions}, Energy and Buildings 253 (2021) 111504.
\newblock \href
  {http://dx.doi.org/https://doi.org/10.1016/j.enbuild.2021.111504}
  {\path{doi:https://doi.org/10.1016/j.enbuild.2021.111504}}.
\newline\urlprefix\url{https://www.sciencedirect.com/science/article/pii/S037877882100788X}

\bibitem{idelchik1986}
I.~E. Idelchik, Hanbook of hydraulic resistance (2nd Edition), Hemisphere
  Publishing Corporation, 1986, {ISBN:} 0-89116-284-4.

\bibitem{ZhuThesis}
Z.~Weimin, \href{https://spectrum.library.concordia.ca/106/}{Characteristics of
  dividing and combining flows}, Ph.d. thesis, Concordia University,
  supervisor: Ramamurthy A. S (1995).
\newline\urlprefix\url{https://spectrum.library.concordia.ca/106/}

\bibitem{miller1990internal}
M.~D.S., B.~(Association),
  \href{https://books.google.es/books?id=5A9nQgAACAAJ}{Internal Flow Systems},
  Miller Innovations, 1990.
\newline\urlprefix\url{https://books.google.es/books?id=5A9nQgAACAAJ}

\bibitem{ZIGANSHIN2021107491}
A.~Ziganshin, K.~Logachev, K.~Batrova,
  \href{https://www.sciencedirect.com/science/article/pii/S0360132320308581}{Reducing
  the drag of midpoint lateral orifices of exhaust air ducts by shaping them
  along vortex zone outlines}, Building and Environment 188 (2021) 107491.
\newblock \href
  {http://dx.doi.org/https://doi.org/10.1016/j.buildenv.2020.107491}
  {\path{doi:https://doi.org/10.1016/j.buildenv.2020.107491}}.
\newline\urlprefix\url{https://www.sciencedirect.com/science/article/pii/S0360132320308581}

\bibitem{TONG2019134}
L.~Tong, J.~Gao, Z.~Luo, L.~Wu, L.~Zeng, G.~Liu, Y.~Wang,
  \href{https://www.sciencedirect.com/science/article/pii/S036013231830742X}{A
  novel flow-guide device for uniform exhaust in a central air exhaust
  ventilation system}, Building and Environment 149 (2019) 134--145.
\newblock \href
  {http://dx.doi.org/https://doi.org/10.1016/j.buildenv.2018.12.007}
  {\path{doi:https://doi.org/10.1016/j.buildenv.2018.12.007}}.
\newline\urlprefix\url{https://www.sciencedirect.com/science/article/pii/S036013231830742X}

\end{thebibliography}

\end{document}